\documentclass[10pt,a4paper]{article}
\usepackage[numbers,sort&compress]{natbib}
\usepackage{authblk}
\usepackage{graphicx}
\usepackage[T1]{fontenc}
\usepackage{lmodern}
\usepackage[top=30truemm,bottom=30truemm,left=25truemm,right=25truemm]{geometry}
\usepackage[colorlinks=true,linktoc=page,citecolor=red,linkcolor=blue]{hyperref}
\graphicspath{{./figs/}}
\usepackage{color,comment}
\usepackage{amsmath,amssymb,amsfonts}
\usepackage{bm,braket,array}
\usepackage[all]{xy}
\usepackage{amsthm,amscd}
\usepackage{slashed}
\usepackage[format=hang,justification=raggedright]{caption}
\usepackage{multirow}
\allowdisplaybreaks[1]
\usepackage{rotating}

\theoremstyle{definition}

\theoremstyle{remark}

\newcommand{\pf}{{\rm Pf\,}}
\newcommand{\Tr}{{\rm Tr\,}}
\newcommand{\tr}{{\rm tr\,}}

\newcommand{\sgn}{{\rm sgn\,}}

\newcommand{\s}{{\sigma}}
\newcommand{\g}{{\gamma}}

\newcommand{\G}{{\Gamma}}

\newcommand{\R}{\mathbb{R}}
\newcommand{\Z}{\mathbb{Z}}

\newcommand{\bx}{{\bm{x}}}
\newcommand{\bk}{{\bm{k}}}

\newcommand{\bR}{{\bm{R}}}
\newcommand{\bG}{{\bm{G}}}

\hypersetup{pdftitle={Equivariant SW},pdfauthor={Ken Shiozaki}}
\begin{document}
\title{A discrete Stiefel–Whitney invariant with twofold rotation symmetry}
\author{Ken Shiozaki}
\affil{Center for Gravitational Physics and Quantum Information, Yukawa Institute for Theoretical Physics, Kyoto University, Kyoto 606-8502, Japan}
\date{\today}
\maketitle

\begin{abstract}
We review Stiefel--Whitney invariants and give a discrete formulation of an additional $\Z_2$ invariant for two-dimensional spinful insulators with twofold rotation and time-reversal symmetries (layer group $p1121'$). A singular gauge transformation specifies a real bundle over the quotient of the Brillouin torus, whose second Stiefel--Whitney number defines the invariant. We derive its dependence on the choice of auxiliary gauge function and prove invariance under changes of Bloch frame and additivity under direct sums. The discrete formula uses overlaps of independently chosen Bloch frames and symmetry-compatible $\mathrm{Pin}_+$ lifts on the boundary of a half Brillouin zone. Atomic and topological-insulator models illustrate the formula and distinguish the additional invariant from the Kane--Mele index. Together with three quantized $\Z_2$ partial polarizations and the filling number, it labels the known stable classification $\Z\oplus\Z_2^{\oplus4}$.
\end{abstract}

\section{Introduction}
\label{sec:introduction}

An antiunitary symmetry that leaves momentum invariant and squares
to $+1$ endows the occupied states with a real vector bundle structure.
Early studies established $\Z_2$ topology in $C_2T$-protected
crystalline insulators~\cite{Ryu_2010,ShiozakiTopology2014,FangFu2015} and in $PT$-protected
nodal-line semimetals~\cite{FangNodal2015}.
A real-bundle formulation was also developed for $PT$-symmetric
Dirac semimetals~\cite{ZhaoReal2017}.
The $\Z_2$ monopole charge of nodal lines was subsequently
identified with the second Stiefel--Whitney (SW) class
~\cite{AhnBandTopology2018}.
SW classes provide a systematic description of topological
obstructions in real bundles
~\cite{AhnSymmetry2019,AhnSWReview2019}.
The relation between the Euler and SW classes for two occupied
bands, as well as fragile topology~\cite{PoFragile2018} and Wannier obstructions,
has also been studied~\cite{AhnEuler2019}.
For related homotopy classifications of $C_2T$-symmetric band structures,
see also Refs.~\cite{PhysRevB.102.115135,PhysRevB.108.155137}.

In a two-dimensional spinful system with separate twofold rotation
$C_2$ and time-reversal $T$ symmetries, the combined symmetry satisfies
$(C_2T)^2=+1$.
The ordinary SW classes, however, do not retain all the information
carried by the separate $C_2$ action.

This loss of information is visible even among atomic insulators in
the layer group $p1121'$.
Consider two occupied Kramers pairs centered at $(0,0)$ and
$(1/2,1/2)$, and compare them with two pairs centered at $(1/2,0)$
and $(0,1/2)$, with positions measured in lattice units relative to
the same unit-cell origin.
These configurations have the same filling, rotation representations
at high-symmetry momenta, and quantized partial polarizations.
Their Kane--Mele index~\cite{Kane22005,FuTime2006} and ordinary first and second SW numbers
vanish in both cases.
Nevertheless, they represent distinct crystalline phases.
This distinction was identified using nested partial polarizations
in Ref.~\cite{KooiClassification2019}; we revisit it in
Sec.~\ref{sec:atomic_c2}.
Thus neither symmetry eigenvalues nor these conventional topological
indices suffice to distinguish atomic insulators.

The stable classification of spinful $C_2$- and $T$-symmetric
insulators is $\Z\oplus\Z_2^{\oplus4}$~\cite{ShiozakiTopology2014,Li_Wan_HomotopyInvariant_C2T}.
The integer counts occupied Kramers pairs.
Three of the $\Z_2$ indices can be expressed as Fu--Kane partial
polarizations quantized by $C_2$ symmetry, with the Kane--Mele index
given by a combination of them~\cite{LauMirror2016,KooiClassification2019}.
The remaining $\Z_2$ component is associated with the two-dimensional
cells in the Atiyah--Hirzebruch spectral sequence~\cite{Shiozaki_AHSS,Li_Wan_HomotopyInvariant_C2T}.

Several works have addressed this remaining crystalline information.
Kruthoff, de Boer, and van Wezel proposed a description in terms of
vortex configurations with identical line and Kane--Mele index~\cite{KruthoffTopology2019}.
Kooi, van Miert, and Ortix used nested partial polarizations to
distinguish atomic configurations with separated Wannier sectors~\cite{KooiClassification2019}.
Related approaches were developed in
Refs.~\cite{KooiHybrid2020,KooiBulkCorner2021}.
Li and Wan constructed a homotopy invariant from Wilson loops
lifted to the Spin group~\cite{Li_Wan_HomotopyInvariant_C2T}.

In a different symmetry setting, Zhang and Zhao studied symmetric
clutching functions and half-BZ invariants for spinless systems with
separate inversion and time-reversal symmetries~\cite{ZhangComplete2026}.
The construction below concerns the spinful case
$C_2^2=T^2=-1$.

In this work, we revisit the remaining $\Z_2$ component through a
systematic $C_2$-equivariant refinement of the SW construction for
spinful $p1121'$ insulators.
We impose $C_2$ symmetry on local real frames and the lifts of their
transition functions.
A singular gauge transformation specified by an auxiliary angular
function $\alpha_\bk$ then defines a real bundle over the quotient
$T^2/C_2\simeq S^2$ of the BZ.
Its second SW number gives $\nu_2^{C_2}[\alpha]$.
We prove invariance under changes of Bloch frame and additivity
under direct sums.
The choice of singular gauge transformation introduces a separate
dependence: changing $\alpha_\bk$ can shift the index by a combination
of the one-dimensional invariants.
We derive this transformation law explicitly.
In particular, the difference between two phases with identical
one-dimensional indices is independent of this choice.

For numerical evaluation, we extend the discrete SW formula of
Ref.~\cite{ShiozakiDiscrete2024} to incorporate the $C_2$ symmetry.
The input consists of occupied Bloch frames chosen independently
at discrete momenta.
After pointwise gauge fixing, including the symmetry conditions at
high-symmetry points, neighboring-frame overlaps determine orthogonal links.
The chosen $\alpha_\bk$ is encoded in symmetry-compatible
$\mathrm{Pin}_+$ lifts on the boundary of a half BZ.
The resulting sum of plaquette signs agrees with the continuous
definition on sufficiently fine meshes.
Atomic and topological-insulator models demonstrate the calculation
and the crystalline distinction that the conventional indices miss.

The paper is organized as follows.
Section~\ref{sec:sw_review} reviews SW classes, their role in band
theory, and the discrete formula for the ordinary second SW number.
Section~\ref{sec:spinful_c2t} treats spinful $p1121'$ insulators:
we review the one-dimensional invariants, construct the
$C_2$-symmetric invariant on the sphere and the torus, and present
atomic examples and the discrete formulation with numerical results.
Section~\ref{sec:conclusion} concludes the paper.
Appendix~\ref{app:alpha_change} derives the dependence on the
singular gauge transformation.

\section{Stiefel--Whitney classes and band theory}
\label{sec:sw_review}
This section reviews the ordinary first and second SW classes, a correction that makes the invariant additive under direct sums, and a discrete formula for the second SW number.

Consider a two-dimensional tight-binding model with lattice constant $1$.
Fix the unit-cell origin, and denote the orbital positions by $\bx_\alpha$,
the internal degrees of freedom by $l$, and the unit cells by $\bR\in\Z^2$.
From the real-space basis $\ket{\bR+\bx_\alpha,l}$, define the momentum-space basis as
\begin{align}
\ket{\bk,\alpha,l}
=\sum_{\bR\in\Z^2}\ket{\bR+\bx_\alpha,l}
e^{i\bk\cdot(\bR+\bx_\alpha)}.
\end{align}
For a reciprocal lattice vector $\bG\in2\pi\Z^2$, this basis satisfies
\begin{align}
\ket{\bk+\bG,\alpha,l}
=\ket{\bk,\alpha,l}e^{i\bG\cdot\bx_\alpha}.
\end{align}
The single-particle Hamiltonian therefore obeys the twisted boundary condition
\begin{align}
H_{\bk+\bG}=V_\bG H_\bk V_\bG^\dagger,\qquad
[V_\bk]_{\alpha l,\alpha'l'}
=e^{-i\bk\cdot\bx_\alpha}\delta_{\alpha\alpha'}\delta_{ll'}.
\end{align}
We distinguish $\bk\in\R^2$ and $\bk+\bG$. 

Assume an antiunitary symmetry that leaves momentum invariant and squares to $+1$.
Examples include $C_2T$ symmetry in spinless or spinful systems and $PT$ symmetry
in spinless systems. In this section, we denote this symmetry by $C_2T$
regardless of its realization, and use a momentum-independent unitary matrix
$D(C_2T)$ satisfying
\begin{align}
&D(C_2T)H_\bk^*D(C_2T)^\dagger=H_\bk,\qquad
D(C_2T)D(C_2T)^*=1_N,
\label{eq:C2Tsym}\\
&D(C_2T)V_\bG^*=V_\bG D(C_2T), 
\label{eq:C2T_VG}
\end{align}
where $N$ is the matrix size of the Hamiltonian $H_\bk$. 
We focus on a set $I$ of $n$ bands separated from the remaining bands by an energy gap.
For the eigenvalues $E_{i\bk}$, we assume that a constant $\Delta E>0$ exists such that
$|E_{i\bk}-E_{j\bk}|>\Delta E$ for all $\bk$, $i\in I$, and $j\notin I$.
An $N\times n$ matrix whose columns form an orthonormal basis of this subspace,
\begin{align}
\Phi_\bk=(\phi_{1\bk},\ldots,\phi_{n\bk}),\qquad
\Phi_\bk^\dagger\Phi_\bk=1_n.
\end{align}
is called a Bloch frame. Its columns may be chosen as eigenstates, but we allow
unitary changes of basis within $I$.

\subsection{Construction of the Stiefel--Whitney classes}
\label{sec:sw_definition}
Choose a good open cover $\{U_a\}_a$ of the BZ torus $T^2$, meaning that every
nonempty finite intersection is contractible. On each patch, choose a smooth
Bloch frame $\Phi^a_\bk$. Across the BZ boundary, extend it according to
\begin{align}
V_\bG\Phi^a_\bk=\Phi^a_{\bk+\bG},\qquad\bk\in U_a.
\end{align}
By the symmetry \eqref{eq:C2Tsym}, each patch admits a $C_2T$ gauge
\begin{align}
D(C_2T)(\Phi^a_\bk)^*=\Phi^a_\bk,
\qquad \bk\in U_a.
\label{eq:C2Tgauge}
\end{align}
On a double overlap $U_{ab}:=U_a\cap U_b$, define the transition function
\begin{align}
t^{ab}_\bk:=(\Phi^a_\bk)^\dagger\Phi^b_\bk\in O(n),
\qquad \bk\in U_{ab}.
\end{align}
It is real and orthogonal by \eqref{eq:C2Tgauge}, and satisfies
\begin{align}
\Phi^b_\bk=\Phi^a_\bk t^{ab}_\bk,\qquad
t^{ba}_\bk=(t^{ab}_\bk)^\top.
\end{align}
On a triple overlap $U_{abc}:=U_a\cap U_b\cap U_c$, it also satisfies
\begin{align}
t^{ab}_\bk t^{bc}_\bk=t^{ac}_\bk.
\end{align}
These transition functions define a real vector bundle of rank $n$ and its
principal $O(n)$ bundle of orthonormal frames. We also denote this real bundle by $\Phi$.

The determinants of the transition functions define a $\Z_2$-valued 1-cocycle through
\begin{align}
\det t^{ab}_\bk=(-1)^{p^{ab}},\qquad p^{ab}\in\{0,1\}.
\end{align}
The first SW class is its cohomology class,
\begin{align}
w_1(\Phi):=[p]\in H^1(T^2,\Z_2).
\end{align}

Next, choose a continuous ${\rm Pin}_+(n)$ lift on each double overlap,
\begin{align}
t^{ab}_\bk\in O(n)
\quad\longmapsto\quad
u(t^{ab}_\bk)\in{\rm Pin}_+(n)
\end{align}
and set
\begin{align}
u(t^{ba}_\bk)=u(t^{ab}_\bk)^\dagger
\end{align}
for the reversed order. Throughout this paper, we use Hermitian Clifford matrices
\begin{align}
\g_a^\dagger=\g_a,\qquad\g_a\g_b+\g_b\g_a=2\delta_{ab},
\label{eq:clifford_convention}
\end{align}
and lift a reflection with unit normal $\bm n$ to $\sum_a n_a\g_a$.
Its square is $+1$, which fixes our ${\rm Pin}_+$ convention.
Note that the notation $u$ denotes a choice of lift, not a global inverse of the covering map.
On a triple overlap,
\begin{align}
u(t^{ab}_\bk)u(t^{bc}_\bk)u(t^{ca}_\bk)
=(-1)^{z^{abc}},\qquad z^{abc}\in\Z_2,
\end{align}
defines a 2-cocycle. The second SW class is
\begin{align}
w_2(\Phi):=[z]\in H^2(T^2,\Z_2).
\end{align}
Changing the local frames or the signs of the lifts may change these cocycles,
but leaves their cohomology classes $w_1,w_2$ unchanged.

\subsection{Whitney sum formula and a corrected invariant}
For another real bundle $\Psi$ of rank $m$, the Whitney sum formula gives
\begin{align}
w_1(\Phi\oplus\Psi)&=w_1(\Phi)+w_1(\Psi),\\
w_2(\Phi\oplus\Psi)&=w_2(\Phi)+w_2(\Psi)
+w_1(\Phi)\cup w_1(\Psi).
\label{eq:Wsumrule2}
\end{align}
The cross term prevents the second SW class from being additive under direct sums
in general. We now introduce a correction that cancels this term on $T^2$.

Abbreviate the generators of $H^1(T^2,\Z_2)$ dual to the fundamental loops in the
$k_x,k_y$ directions as $dk_x,dk_y$, and write
\begin{align}
w_1(\Phi)
=\nu_{1x}(\Phi)dk_x+\nu_{1y}(\Phi)dk_y,\qquad
\nu_{1x},\nu_{1y}\in\{0,1\}.
\end{align}
Here $dk_x,dk_y$ denote normalized cohomology classes, not differential forms.
The first SW numbers $\nu_{1x},\nu_{1y}$ equal the Berry phases in the corresponding
directions divided by $\pi$\footnote{
The Berry phase includes the twist at the BZ boundary. Along
$\ell=\{\bk_0+t\bG\mid0\leq t\leq1\}$, it is defined by
\begin{align}
e^{i\gamma_\bG}
&:=\det\left[\Phi_{\bk_0}^\dagger V_\bG^\dagger\Phi_{\bk_0+\bG}\right]
\lim_{{\cal N}\to\infty}\prod_{j=0}^{{\cal N}-1}
\det\left[\Phi_{\bk_0+\frac{j+1}{\cal N}\bG}^\dagger
\Phi_{\bk_0+\frac{j}{\cal N}\bG}\right]
\end{align}
Then $\nu_{1x}=\gamma_{(2\pi,0)}/\pi$ and
$\nu_{1y}=\gamma_{(0,2\pi)}/\pi$, with equalities understood modulo $2$.
}.
They therefore measure the sum of Wannier centers in the corresponding direction
in units of half the lattice constant. Similarly, using $dk_xdk_y:=dk_x\cup dk_y$, write
\begin{align}
w_2(\Phi)=\nu_2(\Phi)dk_xdk_y,\qquad \nu_2(\Phi)\in\{0,1\},
\end{align}
and call $\nu_2$ the second SW number. All subsequent equations between
$\Z_2$-valued quantities are understood modulo $2$.

The Whitney sum formula \eqref{eq:Wsumrule2} becomes
\begin{align}
\nu_2(\Phi\oplus\Psi)
&=\nu_2(\Phi)+\nu_2(\Psi)
+\nu_{1x}(\Phi)\nu_{1y}(\Psi)
+\nu_{1y}(\Phi)\nu_{1x}(\Psi).
\end{align}
Define the corrected invariant
\begin{align}
\nu_2'(\Phi):=\nu_2(\Phi)+\nu_{1x}(\Phi)\nu_{1y}(\Phi).
\label{eq:w2_redef}
\end{align}
A similar correction was used in Ref.~\cite{ChenTopological2022}.
Under a direct sum, the cross terms cancel, giving
\begin{align}
\nu_2'(\Phi\oplus\Psi)=\nu_2'(\Phi)+\nu_2'(\Psi).
\end{align}
In addition to these three $\Z_2$ numbers, the band number $\nu_0:=n$ is an invariant.

\subsection{Atomic and fragile insulators}
\label{sec:c2t_model}
With only $C_2T$ symmetry imposed, the rank and the first and second SW classes of
the real bundle give the stable classification $\Z\oplus\Z_2^{\oplus3}$
~\cite{ShiozakiTopology2014,AhnSWReview2019}.
It is shown that atomic insulators generate this $K$ group: 
At a generic position in real space, the symmetry class is A.
Class A has no stable topological phase in one dimension, while in two dimensions
it is classified by the Chern number~\cite{SchnyderClassification2008}.
A nonzero Chern number, however, is incompatible with $C_2T$ symmetry.
We calculate the indices of atomic insulators and then give an example that has
no atomic realization with a fixed number of bands.

Let ${\mathsf a}_{{\rm x}{\rm y}}$ denote an atomic insulator with one localized orbital at a high-symmetry position of the unit cell,
\begin{align}
({\rm x},{\rm y})\in
\left\{(0,0),\left(\tfrac12,0\right),
\left(0,\tfrac12\right),\left(\tfrac12,\tfrac12\right)\right\}.
\end{align}
The models ${\mathsf a}_{{\rm x}{\rm y}}$ in this section have one band, so $\nu_2=0$.
Their Wannier centers give
\begin{align}
\nu_{1x}({\mathsf a}_{{\rm x}{\rm y}})=2{\rm x},\qquad
\nu_{1y}({\mathsf a}_{{\rm x}{\rm y}})=2{\rm y}.
\end{align}
Applying the correction \eqref{eq:w2_redef} gives Table~\ref{tab:c2t}.

The indices of an atomic insulator are nonnegative integer sums of those of the
four atomic orbitals in the table. If the band number is $2$ and
$\nu_{1x}=\nu_{1y}=0$, the two orbitals must have the same first SW numbers;
additivity then gives $\nu_2'=0$.
Thus an atomic insulator cannot realize
$(\nu_0,\nu_{1x},\nu_{1y},\nu_2')=(2,0,0,1)$.

As a band-insulator realization of these indices, consider four orbitals at the
unit-cell center with
\begin{align}
H_\bk&=\sin k_x\s_x-\sin k_y\s_y\tau_z
+(1-\cos k_x-\cos k_y)\s_z,\\
V_\bG&=1_4,\qquad D(C_2T)=\tau_x.
\label{eq:c2t_FI_model}
\end{align}
Denote its two negative-energy bands by ${\sf BI}$.
Here $\s_\mu,\tau_\mu$ are Pauli matrices acting on two degrees of freedom.
The $\tau_z=\pm1$ sectors of \eqref{eq:c2t_FI_model} carry opposite Chern numbers
$\pm1$ and are exchanged by $C_2T$.
The associated real bundle is the underlying real bundle of the occupied complex
line bundle in one sector. It is therefore orientable, giving
$\nu_{1x}({\sf BI})=\nu_{1y}({\sf BI})=0$~\footnote{
To see the real-bundle structure explicitly, choose a normalized
occupied state $u^a_\bk$ in the $\tau_z=+1$ sector on each patch.
Its $C_2T$ partner lies in the $\tau_z=-1$ sector, and the two columns of
\begin{align}
\Phi^a_\bk
:=\frac{1}{\sqrt{2}}
\left(
u^a_\bk+D(C_2T)(u^a_\bk)^*,\;
i\bigl[u^a_\bk-D(C_2T)(u^a_\bk)^*\bigr]
\right)
\end{align}
form an orthonormal frame satisfying
$D(C_2T)(\Phi^a_\bk)^*=\Phi^a_\bk$.
On a patch overlap, a phase change
$u^b_\bk=e^{i\theta^{ab}_\bk}u^a_\bk$ induces
\begin{align}
\Phi^b_\bk
=\Phi^a_\bk
\begin{pmatrix}
\cos\theta^{ab}_\bk & -\sin\theta^{ab}_\bk\\
\sin\theta^{ab}_\bk & \cos\theta^{ab}_\bk
\end{pmatrix}.
\end{align}
This identifies the real occupied bundle with the underlying
real bundle of the occupied complex line bundle in one sector.
Since all transition matrices have determinant $+1$, this bundle
is orientable.}.
Its second SW number is the Chern number reduced modulo $2$~\footnote{
The relation between the second SW number and the Chern number
can be seen from the transition functions introduced in the previous footnote. 
On a triple overlap, the phases satisfy $\theta^{ab}_\bk+\theta^{bc}_\bk-\theta^{ac}_\bk
=2\pi n^{abc}, n^{abc}\in\Z$.
These integers form the cocycle representing the first Chern class
of the occupied complex line bundle.
The corresponding $\mathrm{Spin}(2)$ lifts of the real transition
matrices can be chosen as $u(t^{ab}_\bk)
=e^{-\frac{\theta^{ab}_\bk}{2}\g_1\g_2}$. 
Since $(\g_1\g_2)^2=-1$, their product on a triple overlap is $u(t^{ab}_\bk)u(t^{bc}_\bk)u(t^{ac}_\bk)^\dagger
=e^{-\pi n^{abc}\g_1\g_2}
=(-1)^{n^{abc}}$. 
Thus the cocycle defining $w_2$ is precisely the Chern cocycle
reduced modulo $2$.}, so $\nu_2({\sf BI})=\nu_2'({\sf BI})=1$~\cite{AhnSWReview2019,AhnEuler2019}.

A phase whose Wannier obstruction is removed by adding atomic bands is called
a fragile insulator~\cite{PoFragile2018}.
For this example, the relation in the $K$ group is
\begin{align}
{\sf BI}
\sim {\mathsf a}_{\frac12\frac12}\oplus{\mathsf a}_{\frac120}
\oplus{\mathsf a}_{0\frac12}\ominus{\mathsf a}_{00}
\end{align}
expressing the occupied bands as a formal difference of atomic bands.

\begin{table}[tb]
\centering
\caption{Indices of atomic and band insulators with $C_2T$ symmetry.
$\nu_0$ is the band number, $\nu_{1x},\nu_{1y}$ are the first SW numbers,
and $\nu_2'$ is the additive $\Z_2$ invariant defined in \eqref{eq:w2_redef}.
Each atomic insulator in this table consists of one band.}
\label{tab:c2t}
\begin{tabular}{c|cccc}
&$\nu_0$&$\nu_{1x}$&$\nu_{1y}$&$\nu_2'$\\\hline
${\mathsf a}_{00}$&1&0&0&0\\
${\mathsf a}_{\frac120}$&1&1&0&0\\
${\mathsf a}_{0\frac12}$&1&0&1&0\\
${\mathsf a}_{\frac12\frac12}$&1&1&1&1\\
${\sf BI}$&2&0&0&1
\end{tabular}
\end{table}

\subsection{A discrete formula for the second SW number}
\label{sec:dis_w2}
We present the discrete formula for the second SW number $\nu_2$
~\cite{ShiozakiDiscrete2024}, including twisted boundary conditions.
Section~\ref{sec:w2_c2} extends it to systems with additional $C_2$ symmetry.

Discretize momentum space on a square mesh, and let $\hat\mu$ denote one mesh step
in the $\mu=x,y$ direction. Choose a Bloch frame at each vertex satisfying
\begin{align}
V_\bG\Phi_\bk&=\Phi_{\bk+\bG},
\label{eq:lat_pgauge}\\
D(C_2T)(\Phi_\bk)^*&=\Phi_\bk.
\label{eq:lat_c2t_gauge}
\end{align}
The frames need not be chosen smoothly between neighboring vertices.
The gauge \eqref{eq:lat_c2t_gauge} can be obtained from the Takagi factorization
$D(C_2T)=QQ^\top$, $Q\in U(N)$.
Indeed, \eqref{eq:C2Tsym} makes $Q^\dagger H_\bk Q$ real and symmetric;
multiplying its real orthonormal eigenvectors by $Q$ gives the desired frame.

For each edge $(\bk,\bk+\hat\mu)$, take the singular value decomposition of the
overlap matrix and define an orthogonal link:
\begin{align}
\Phi_\bk^\dagger\Phi_{\bk+\hat\mu}= L\Sigma  R^\top
\quad\longmapsto\quad
t^\mu_\bk:=L R^\top\in O(n).
\end{align}
We assume that the overlap matrix is nonsingular.
The boundary condition \eqref{eq:lat_pgauge} implies
\begin{align}
t^\mu_{\bk+\bG}=t^\mu_\bk.
\end{align}
Choose a lift on each edge and extend it periodically:
\begin{align}
t^\mu_\bk\longmapsto u(t^\mu_\bk)\in{\rm Pin}_+(n),\qquad
u(t^\mu_{\bk+\bG})=u(t^\mu_\bk).
\end{align}
For a numerical construction of the lifts, see Ref.~\cite{ShiozakiDiscrete2024}.

On a sufficiently fine mesh, the product of lifts around a plaquette
$\Box_\bk=(\bk,\bk+\hat x,\bk+\hat x+\hat y,\bk+\hat y)$
is close to either $+1$ or $-1$. Define its $\Z_2$ value by this sign:
\begin{align}
u(t^x_\bk)u(t^y_{\bk+\hat x})
u(t^x_{\bk+\hat y})^\dagger u(t^y_\bk)^\dagger
\sim(-1)^{z_{\Box_\bk}}1,\qquad z_{\Box_\bk}\in\{0,1\}.
\end{align}
Numerically, we use the sign of the real part of the trace of the left-hand side
and check that it is nonzero. The second SW number is the sum over the entire BZ,
\begin{align}
\nu_2=\sum_{\Box_\bk\in T^2}z_{\Box_\bk}\quad\bmod2.
\end{align}
Changing the sign of an edge lift changes both adjacent plaquette signs and
therefore leaves this sum unchanged.

\section[Layer group p1121' for spinful electrons]{Layer group $p1121'$ for spinful electrons}
\label{sec:spinful_c2t}

We consider a two-dimensional spinful electron system with layer
group $p1121'$ symmetry.
The crystalline $\Z_2$ classification component beyond the
one-dimensional indices has been studied from several
perspectives~\cite{KruthoffTopology2019,KooiClassification2019,
KooiHybrid2020,Li_Wan_HomotopyInvariant_C2T}.
Here we revisit this component through a $C_2$-symmetric
Stiefel--Whitney construction and develop its discrete formulation.

In addition to Eqs.~\eqref{eq:C2Tsym} and \eqref{eq:C2T_VG}, the symmetry conditions are
\begin{align}
D(C_2)H_\bk D(C_2)^\dagger&=H_{-\bk},
\quad D(C_2)^2=-1_N,\notag\\
D(C_2)V_\bG&=V_{-\bG}D(C_2),\notag\\
D(C_2)D(C_2T)&=D(C_2T)D(C_2)^*,
\end{align}
where $D(C_2)$ is a momentum-independent unitary matrix.
At the high-symmetry points
\begin{align}
\G=(0,0),\qquad X=(\pi,0),\qquad Y=(0,\pi),\qquad M=(\pi,\pi),
\end{align}
we have
\begin{align}
V_{2P}D(C_2)H_P\bigl(V_{2P}D(C_2)\bigr)^\dagger&=H_P,\notag\\
D(C_2T)H_P^*D(C_2T)^\dagger&=H_P.
\end{align}
Furthermore,
\begin{align}
\bigl(V_{2P}D(C_2)\bigr)^2&=-1_N,\notag\\
\bigl(V_{2P}D(C_2)\bigr)D(C_2T)
&=D(C_2T)\bigl(V_{2P}D(C_2)\bigr)^*
\end{align}
imply that the antiunitary $C_2T$ operation exchanges the sectors with rotation eigenvalues $+i,-i$.
Corresponding states in the two sectors have the same energy and form Kramers pairs under time reversal.

The occupied band number is therefore even. We write $\nu_0=2m$ and define the standard antisymmetric matrix on the occupied space by
\begin{align}
J:=\bigoplus_{a=1}^{m}i\s_y,
\qquad J^\top=-J,\quad J^2=-1_{2m}.
\label{eq:standard_J}
\end{align}
Throughout, $\s_x,\s_y,\s_z$ denote the $2\times2$ Pauli matrices.

The stable classification is
\begin{align}
K\cong\Z\oplus\Z_2^{\oplus4}
\end{align}
where the integer component counts Kramers pairs, $\nu_0/2$
~\cite{ShiozakiTopology2014}.
In the Atiyah--Hirzebruch spectral sequence~\cite{Shiozaki_AHSS}, three of the four $\Z_2$ components arise from the 1-skeleton and one from the 2-cell
~\cite{Li_Wan_HomotopyInvariant_C2T}.

Their relation to the ordinary SW numbers requires care.
The occupied rotation eigenvalues at each high-symmetry point occur in $+i,-i$ pairs,
so the total Berry phase vanishes along each of the two fundamental loops, giving
\begin{align}
\nu_{1x}=\nu_{1y}=0.
\end{align}
The ordinary second SW number, however, equals the Kane--Mele invariant and need not vanish
~\cite{Li_Wan_HomotopyInvariant_C2T}:
\begin{align}
\nu_2=\nu_2'=\nu_{\rm KM}\pmod2.
\label{eq:ordinary_sw_km}
\end{align}
Since the trace of the Berry curvature vanishes due to $C_2T$ symmetry, the Kane--Mele invariant is determined by the indices on the 1-skeleton
and must be distinguished from the additional invariant associated with the 2-cell.

\subsection{Invariants on one-dimensional subspaces}
\label{sec:partial_polarization}

We define invariants on one-dimensional subspaces using the Fu--Kane partial polarization
~\cite{FuTime2006}.
Let the matrix representing time reversal be
\begin{align}
D(T):=D(C_2)D(C_2T),\qquad D(T)D(T)^*=-1_N.
\end{align}
At a high-symmetry point $P$, the sewing matrix of the occupied frame $\Phi_P$,
$\Phi_P^\dagger V_{2P}D(T)\Phi_P^*$, is antisymmetric and unitary, so its Pfaffian is defined.

Consider the four intervals $(P,Q)=(\G,X),(Y,M),(\G,Y),(X,M)$.
We denote the partial polarization multiplied by $2\pi$ by
$\g_{\rm T}(P\to Q)\in\R/2\pi\Z$ and define
\begin{align}
e^{i\g_{\rm T}(P\to Q)}
:={}&\frac{\pf\!\left(\Phi_P^\dagger V_{2P}D(T)\Phi_P^*\right)}
{\pf\!\left(\Phi_Q^\dagger V_{2Q}D(T)\Phi_Q^*\right)}\notag\\
&\times\lim_{{\cal N}\to\infty}
\prod_{j=0}^{{\cal N}-1}
\det\!\left[\Phi_{P+(j+1)\bm\delta}^\dagger\Phi_{P+j\bm\delta}\right],
\qquad\bm\delta=\frac{Q-P}{{\cal N}}.
\label{eq:partial_polarization}
\end{align}
The endpoint determinant factors from a gauge change cancel between the Pfaffian ratio and the overlap product, making this phase gauge invariant.
At the BZ boundary, we use the boundary conditions specified by $V_\bG$.

Regarding $H_{k_x,0}$ as a one-dimensional system,
$\g_{\rm T}(\G\to X)/(2\pi)$ is the sum of the Wannier centers obtained by choosing one state from each Kramers pair, modulo the lattice period.
Positions are measured from the unit-cell center. The other three intervals have the same interpretation.

We use the following conventions for the Berry connection and curvature:
\begin{align}
A_\mu&=i\Phi_\bk^\dagger\partial_{k_\mu}\Phi_\bk,\notag\\
F_\bk&=\partial_{k_x} A_y-\partial_{k_y} A_x
-i[ A_x, A_y].
\label{eq:berry_convention}
\end{align}
Orienting the half BZ $[-\pi,\pi]\times[0,\pi]$ by $dk_x\wedge dk_y$, we have
\begin{align}
\nu_{\rm KM}
=\frac{\g_{\rm T}(\G\to X)-\g_{\rm T}(Y\to M)}{\pi}
-\frac{1}{2\pi}\int_{\frac12T^2}\tr F_\bk\,d^2k
\pmod2
\end{align}
for a general time-reversal-symmetric system~\cite{FuTime2006}.

With the additional $C_2$ symmetry, the phase is quantized as~\cite{LauMirror2016}
\begin{align}
\g_{\rm T}(P\to Q)\in\{0,\pi\}\pmod{2\pi}.
\end{align}
Indeed, in a smooth $C_2T$-real gauge along the interval, the endpoint sewing matrices are real antisymmetric orthogonal matrices and have Pfaffians $\pm1$.
The determinant of the overlap product is also real, proving the quantization.
In terms of Wannier centers, this means that position inversion under $C_2$ preserves the partial polarization.
Individual centers need not lie at high-symmetry positions: Kramers pairs at general positions may also be exchanged by $C_2$.

The $C_2T$ symmetry also implies $\tr F_\bk=0$.
Joining the four intervals in the order $\G\to X\to M\to Y\to\G$ cancels the endpoint Pfaffians, and Stokes' theorem gives
\begin{align}
\g_{\rm T}(\G\to X)-\g_{\rm T}(Y\to M)
\equiv\g_{\rm T}(\G\to Y)-\g_{\rm T}(X\to M)
\pmod{2\pi}.
\label{eq:gT_rel}
\end{align}
We may therefore choose the three independent indices~\cite{Li_Wan_HomotopyInvariant_C2T}:
\begin{align}
\nu^{C_2}_{1,P\to Q}:=\frac{\g_{\rm T}(P\to Q)}{\pi}\pmod2,
\qquad(P,Q)=(\G,X),(Y,M),(\G,Y).
\label{eq:partial_indices}
\end{align}
Since Pfaffians and determinants are multiplicative under direct sums,
\begin{align}
\nu^{C_2}_{1,P\to Q}(\Phi\oplus\Psi)
=\nu^{C_2}_{1,P\to Q}(\Phi)+\nu^{C_2}_{1,P\to Q}(\Psi)\pmod2.
\end{align}
Moreover, the curvature integral vanishes, yielding
\begin{align}
\nu_{\rm KM}
=\nu^{C_2}_{1,\G\to X}-\nu^{C_2}_{1,Y\to M}\pmod2.
\label{eq:km_partial_indices}
\end{align}
Thus, the Kane--Mele invariant is not an additional independent index.

\subsection[C2-symmetry-enriched Z2 invariant on the sphere S2]{$C_2$-symmetry-enriched $\Z_2$ invariant on the sphere $S^2$}
\label{sec:case_s2}

Before turning to the torus, we consider the sphere $S^2$.
The $C_2$ rotation fixes the north and south poles, ${\rm N}$ and ${\rm S}$,
and acts on spherical coordinates as $(\theta,\phi)\mapsto(\theta,\phi+\pi)$.
Each hemisphere contracts equivariantly to its pole, so for ${\rm P}={\rm N,S}$ we can choose a smooth frame satisfying
\begin{align}
D(C_2T)(\Phi^{\rm P}_\bk)^*&=\Phi^{\rm P}_\bk,\notag\\
D(C_2)\Phi^{\rm P}_\bk&=\Phi^{\rm P}_{C_2\bk}J.
\end{align}
Define the transition function on the equator by
\begin{align}
t^{\rm NS}_\phi:=(\Phi^{\rm N}_\phi)^\dagger\Phi^{\rm S}_\phi\in O(2m).
\end{align}
It satisfies
\begin{align}
t^{\rm NS}_{\phi+\pi}=Jt^{\rm NS}_\phi J^\top
\label{eq:s2_tr_sym}.
\end{align}
Hence
\begin{align}
\tilde t^{\rm NS}_\phi:=e^{-\phi J/2}t^{\rm NS}_\phi e^{\phi J/2}
\label{eq:s2_tilde_t}
\end{align}
is $\pi$-periodic, and so $\tilde t^{\rm NS}_\phi$
defines a loop in $O(2m)$ for $\phi\in[0,\pi]$.
Since
\begin{align}
\pi_1[O(2m)]\cong
\begin{cases}
\Z &(m=1),\\
\Z_2 &(m\geq2),
\end{cases}
\end{align}
we obtain a $\Z_2$ number in the stable range.

This index is independent of symmetry-preserving frame changes.
Under $\Phi^{\rm P}_\bk\mapsto\Phi^{\rm P}_\bk W^{\rm P}_\bk$, we have
\begin{align}
W^{\rm P}_{C_2\bk}=JW^{\rm P}_\bk J^\top,\qquad
\tilde W^{\rm P}_\bk:=e^{-\phi J/2}W^{\rm P}_\bk e^{\phi J/2}.
\end{align}
The matrix $\tilde W^{\rm P}$ is $\pi$-periodic, and its limit at the pole is independent of $\phi$ because $[W^{\rm P},J]=0$ there.
The gauge loop $\tilde W^P_\phi$ on the equator therefore contracts over the quotient disk of the hemisphere to its constant value at the pole.
This constant lies in the connected group $U(m)$ and can be further deformed to the identity.
Consequently,
\begin{align}
\tilde t^{\rm NS}_\phi\longmapsto
(\tilde W^{\rm N}_\phi)^\top\tilde t^{\rm NS}_\phi\tilde W^{\rm S}_\phi
\end{align}
preserves the homotopy class of the loop.

\subsection[C2-symmetry-enriched Z2 invariant on the torus T2]{$C_2$-symmetry-enriched $\Z_2$ invariant on the torus $T^2$}
\label{sec:c2z2top_torus}

Now we construct a real bundle over the quotient space $Q:=T^2/C_2\simeq S^2$
from the transition functions of local frames.

\subsubsection[Definition of the Z2 invariant]{Definition of the $\Z_2$ invariant $\nu_2^{C_2}$}

\begin{figure}[tb]
\centering
\includegraphics[width=0.7\linewidth]{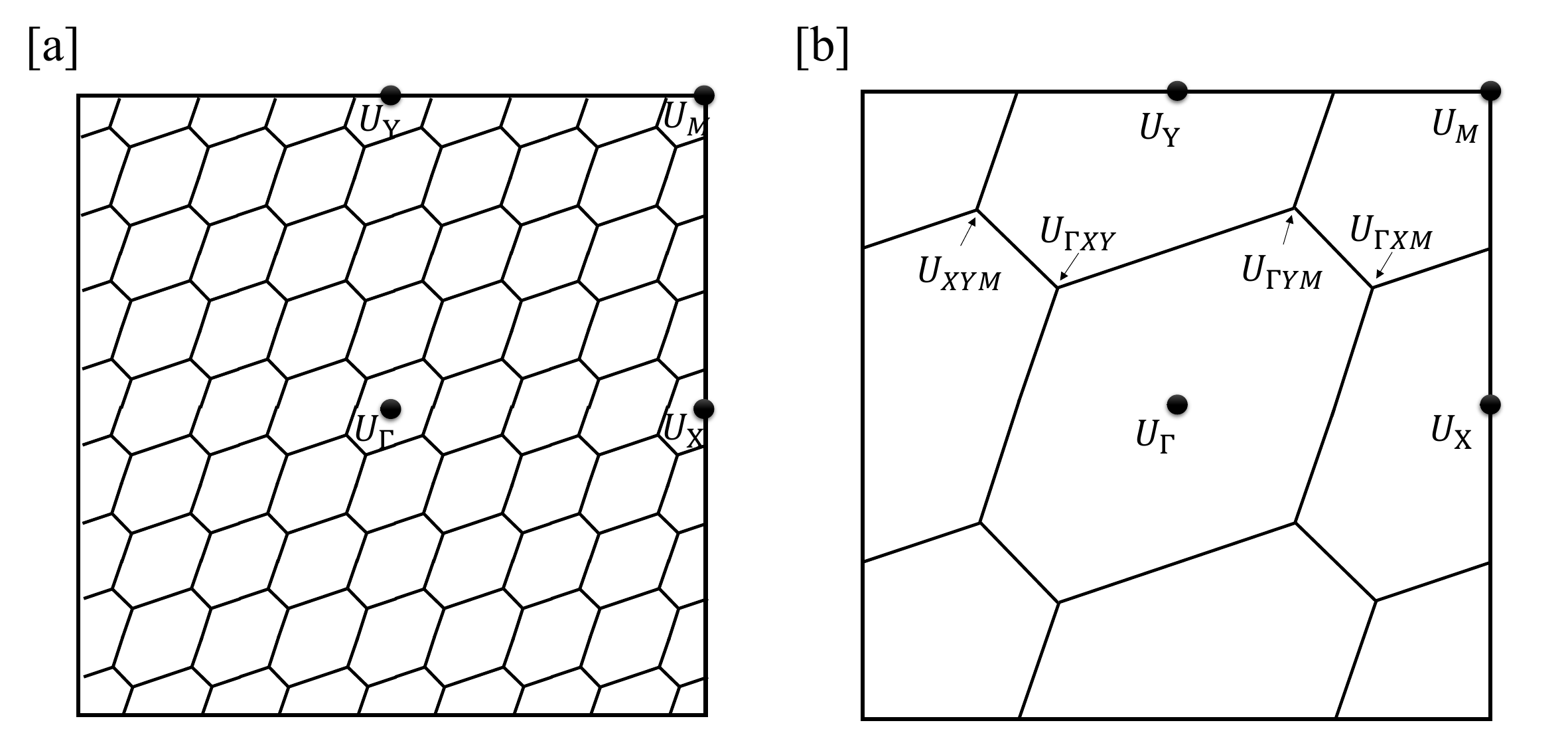}
\caption{A $C_2$-symmetric cover of the BZ torus.
Double and triple overlaps are shown schematically as lines and junctions, respectively.
The cover is chosen so that every nonempty intersection is contractible on the quotient space $Q$.
[b] A cover with four patches, whose independent triple overlaps are
$U_{\G XY},U_{\G XM},U_{\G YM},U_{XYM}$.}
\label{fig:c2patch}
\end{figure}

Pick a $C_2$-symmetric cover $\{U_P\}_P$ of the BZ torus. 
See Fig.~\ref{fig:c2patch} for examples. 
Extend the cover to $\R^2$ by reciprocal-lattice translations and $C_2$ rotations,
using the same label for symmetry-equivalent patches.
On each patch, choose a continuous local frame satisfying
\begin{align}
D(C_2T)(\Phi^P_\bk)^*&=\Phi^P_\bk,\notag\\
D(C_2)\Phi^P_\bk&=\Phi^P_{-\bk}J,\notag\\
V_\bG\Phi^P_\bk&=\Phi^P_{\bk+\bG}.
\label{eq:c2_gauge}
\end{align}
This gauge exists because each patch contracts equivariantly to a fixed point,
where the rotation eigenvalues $+i,-i$ have equal multiplicities.
Define the transition functions by
\begin{align}
t^{PQ}_\bk:=(\Phi^P_\bk)^\dagger\Phi^Q_\bk\in O(2m).
\end{align}
These functions satisfy
\begin{align}
t^{PQ}_{-\bk}=Jt^{PQ}_\bk J^\top,\qquad
t^{PQ}_{\bk+\bG}=t^{PQ}_\bk.
\end{align}
We write the determinant parity as 
\begin{align}
    p^{PQ}=(1-\det t^{PQ})/2 \in \{0,1\}. 
\end{align}
It is constant on each connected component of an overlap.

Imposing the analogous condition on the lifts,
\begin{align}
u(t^{PQ}_{-\bk})=u(J)u(t^{PQ}_\bk)u(J)^\dagger
\label{eq:c2_sym_t}
\end{align}
is impossible when $p^{PQ}=1$. To see this, let
\begin{align}
\Omega:=\g_1\cdots\g_{2m},\qquad u(J)^2=\pm\Omega.
\label{eq:clifford_volume}
\end{align}
The element $\Omega$ lifts $-1_{2m}$ and satisfies
\begin{align}
\Omega u(t^{PQ})\Omega^\dagger=(-1)^{p^{PQ}}u(t^{PQ}).
\label{eq:clifford_parity}
\end{align}
Applying Eq.~\eqref{eq:c2_sym_t} twice would then give
$u(t^{PQ})=(-1)^{p^{PQ}}u(t^{PQ})$, a contradiction for $p^{PQ}=1$.

We therefore choose an angle function $\alpha_\bk\in\R/2\pi\Z$,
continuous away from the high-symmetry points, such that
\begin{align}
\alpha_{-\bk}=\alpha_\bk+\pi,\qquad
\alpha_{\bk+\bG}=\alpha_\bk\pmod{2\pi}.
\label{eq:alpha_cond}
\end{align}
In the examples below, we use
\begin{align}
\alpha^0_\bk&:=\operatorname{Arg}(\sin k_x+i\sin k_y),
\label{eq:canonical_alpha}\\
\alpha^1_\bk&:=\operatorname{Arg}(\sin(k_x+k_y)+i\sin k_y).
\label{eq:alternative_alpha}
\end{align}
Here $\operatorname{Arg}$ is understood as an angle modulo $2\pi$;
we specify a branch range whenever a real-valued angle is needed.

Fixing $\alpha$, we apply the singular gauge transformation
\begin{align}
\tilde\Phi^P_\bk:=\Phi^P_\bk
e^{\frac{\alpha_\bk}{2}(i1_{2m}-J)}.
\label{eq:singular_gauge_tr_c2}
\end{align}
The exponential is $2\pi$-periodic in $\alpha$.
Since the overlaps contain no high-symmetry points, the transition functions
\begin{align}
\tilde t^{PQ}_\bk
:=(\tilde\Phi^P_\bk)^\dagger\tilde\Phi^Q_\bk
=e^{\alpha_\bk J/2}t^{PQ}_\bk e^{-\alpha_\bk J/2}
\label{eq:transition_func_singular_gauge}
\end{align}
are continuous and single-valued on each overlap.
The transformed frames satisfy
\begin{align}
D(C_2T)(\tilde\Phi^P_\bk)^*&=\tilde\Phi^P_\bk e^{-i\alpha_\bk},\notag\\
D(C_2)\tilde\Phi^P_\bk&=i\tilde\Phi^P_{-\bk},\notag\\
V_\bG\tilde\Phi^P_\bk&=\tilde\Phi^P_{\bk+\bG},
\label{eq:c2_tilde_gauge}
\end{align}
and hence
\begin{align}
(\tilde t^{PQ}_\bk)^*=\tilde t^{PQ}_\bk,\qquad
\tilde t^{PQ}_{-\bk}=\tilde t^{PQ}_\bk,\qquad
\tilde t^{PQ}_{\bk+\bG}=\tilde t^{PQ}_\bk.
\end{align}
The real bundle on $Q$ is defined by gluing trivial real bundles
over the quotient patches using the transition functions $\tilde t^{PQ}$.
These functions are continuous on the overlaps, which avoid
the fixed points, and satisfy the cocycle condition.
Thus the bundle is well defined on all of $Q$, without requiring
the transformed Bloch frames to extend to the fixed points. 
We define $\nu_2^{C_2}[\alpha]$ as the second SW number of this bundle.

Choose continuous ${\rm Pin}_+(2m)$ lifts on the double overlaps in the quotient space.
Pulling them back to the torus gives
\begin{align}
u(\tilde t^{PQ}_{-\bk})&=u(\tilde t^{PQ}_\bk),\notag\\
u(\tilde t^{PQ}_{\bk+\bG})&=u(\tilde t^{PQ}_\bk),\notag\\
u(\tilde t^{QP}_\bk)&=u(\tilde t^{PQ}_\bk)^\dagger.
\label{eq:c2sym_lift_cond}
\end{align}
On each triple overlap, define
\begin{align}
(-1)^{z^{PQR}_{\alpha,\bk}}
:=u(\tilde t^{PQ}_\bk)u(\tilde t^{QR}_\bk)u(\tilde t^{RP}_\bk).
\label{eq:def_tilde_z}
\end{align}
This value is constant on each connected component, periodic, and $C_2$-invariant,
so we omit $\bk$ below. For the cover in Fig.~\ref{fig:c2patch}[b],
\begin{align}
\nu_2^{C_2}[\alpha]
=z^{\G XY}_\alpha+z^{\G XM}_\alpha+z^{\G YM}_\alpha+z^{XYM}_\alpha
\quad\bmod2.
\label{eq:c2z2inv_def}
\end{align}

Multiplying the lift on each independent double overlap by $(-1)^{\eta^{PQ}}$ gives
\begin{align}
z^{PQR}_\alpha\longmapsto
z^{PQR}_\alpha+\eta^{PQ}+\eta^{QR}+\eta^{RP}\pmod2.
\end{align}
Each $\eta$ appears twice in the sum over the four triple overlaps.
Thus, $\nu_2^{C_2}[\alpha]$ is independent of the lift signs.

\subsubsection{Gauge invariance}

A frame change $\Phi^P_\bk\mapsto\Phi^P_\bk W^P_\bk$ preserving Eq.~\eqref{eq:c2_gauge}
is given by a continuous orthogonal matrix $W^P_\bk$ satisfying
\begin{align}
W^P_{-\bk}=JW^P_\bk J^\top,\qquad
W^P_{\bk+\bG}=W^P_\bk.
\label{eq:WP_gauge}
\end{align}
After the singular gauge transformation, it acts as
\begin{align}
\tilde\Phi^P_\bk&\longmapsto\tilde\Phi^P_\bk\tilde W^P_\bk,\notag\\
\tilde W^P_\bk&:=e^{\alpha_\bk J/2}W^P_\bk e^{-\alpha_\bk J/2}.
\end{align}
Since $[W^P_P,J]=0$ at a fixed point,
$\lim_{\bk\to P}\tilde W^P_\bk=W^P_P$.
Moreover, $\tilde W^P$ is real orthogonal, periodic, and $C_2$-invariant.
It therefore defines a continuous gauge transformation on the quotient patch and admits a continuous lift.
Choosing the transformed transition functions and their lifts as
\begin{align}
\tilde t^{PQ}_\bk&\longmapsto
(\tilde W^P_\bk)^\top\tilde t^{PQ}_\bk\tilde W^Q_\bk,\notag\\
u(\tilde t^{PQ}_\bk)&\longmapsto
u(\tilde W^P_\bk)^\dagger u(\tilde t^{PQ}_\bk)u(\tilde W^Q_\bk),
\end{align}
the triple product transforms as
\begin{align}
(-1)^{z^{PQR}_\alpha}\longmapsto
u(\tilde W^P)^\dagger(-1)^{z^{PQR}_\alpha}u(\tilde W^P)
=(-1)^{z^{PQR}_\alpha}.
\end{align}
Thus, $\nu_2^{C_2}[\alpha]$ is independent of allowed frame changes.

\subsubsection{Direct-sum rule}

For two occupied bundles $\Phi,\Psi$, use the same cover and $\alpha$,
and arrange the blocks of $J$ to respect the direct sum. Then
\begin{align}
\tilde t^{PQ}_{\Phi\oplus\Psi}
=\tilde t^{PQ}_\Phi\oplus\tilde t^{PQ}_\Psi.
\end{align}
We omit the common $\alpha,\bk$ below.
Order the patches as $\G<X<Y<M$ and embed the lifts for the two systems into the Clifford algebra of the direct-sum space.
For $P<Q$, choose
\begin{align}
u(\tilde t^{PQ}_{\Phi\oplus\Psi})
=u(\tilde t^{PQ}_\Phi)u(\tilde t^{PQ}_\Psi)
\end{align}
and use the Hermitian conjugate for the reverse direction.

Gamma matrices belonging to different systems anticommute.
Since the parity of the number of reflections equals $p^{PQ}_\Phi=(1-\det\tilde t^{PQ}_\Phi)/2$,
\begin{align}
u(\tilde t^{PQ}_\Phi)u(\tilde t^{RS}_\Psi)
=(-1)^{p^{PQ}_\Phi p^{RS}_\Psi}
u(\tilde t^{RS}_\Psi)u(\tilde t^{PQ}_\Phi).
\end{align}
Applying this relation to the product for $P<Q<R$ gives
\begin{align}
&u(\tilde t^{PQ}_{\Phi\oplus\Psi})u(\tilde t^{QR}_{\Phi\oplus\Psi})\notag\\
&=u(\tilde t^{PQ}_\Phi)u(\tilde t^{PQ}_\Psi)
  u(\tilde t^{QR}_\Phi)u(\tilde t^{QR}_\Psi)\notag\\
&=(-1)^{p^{PQ}_\Psi p^{QR}_\Phi}
  u(\tilde t^{PQ}_\Phi)u(\tilde t^{QR}_\Phi)
  u(\tilde t^{PQ}_\Psi)u(\tilde t^{QR}_\Psi)\notag\\
&=(-1)^{z^{PQR}_\Phi+z^{PQR}_\Psi+p^{PQ}_\Psi p^{QR}_\Phi}
  u(\tilde t^{PR}_{\Phi\oplus\Psi}).
\end{align}
Thus,
\begin{align}
z^{PQR}_{\Phi\oplus\Psi}
=z^{PQR}_\Phi+z^{PQR}_\Psi+p^{PQ}_\Psi p^{QR}_\Phi\pmod2.
\end{align}
The sum of the cross terms over the four triple overlaps is
\begin{align}
&p^{\G X}_\Psi p^{XY}_\Phi+p^{\G X}_\Psi p^{XM}_\Phi
+p^{\G Y}_\Psi p^{YM}_\Phi+p^{XY}_\Psi p^{YM}_\Phi\notag\\
&=p^{\G X}_\Psi(p^{XY}_\Phi+p^{XM}_\Phi)
 +(p^{\G Y}_\Psi+p^{XY}_\Psi)p^{YM}_\Phi\notag\\
&=p^{\G X}_\Psi p^{YM}_\Phi+p^{\G X}_\Psi p^{YM}_\Phi=0\pmod2,
\end{align}
where we used the cocycle condition
$p^{PQ}+p^{QR}=p^{PR}\pmod2$ for the transition functions. Therefore,
\begin{align}
\nu_2^{C_2}[\alpha](\Phi\oplus\Psi)
=\nu_2^{C_2}[\alpha](\Phi)+\nu_2^{C_2}[\alpha](\Psi)\pmod2.
\label{eq:c2_whitney_additivity}
\end{align}

\subsubsection{Dependence on the singular gauge transformation}

Let $\beta_\bk:=\alpha'_\bk-\alpha_\bk\in\R/2\pi\Z$ be the difference between two functions $\alpha,\alpha'$ satisfying Eq.~\eqref{eq:alpha_cond}.
Since
\begin{align}
\beta_{-\bk}=\beta_\bk,\qquad\beta_{\bk+\bG}=\beta_\bk\pmod{2\pi},
\label{eq:beta_cond}
\end{align}
$\beta$ defines an angle function on $Q^\circ:=(T^2\setminus F)/C_2=Q\setminus F$, where $F:=\{\G,X,Y,M\}$ is the set of fixed points.
To make the choice explicit, we write the transition functions as $\tilde t^{PQ}_{\alpha,\bk}$ below.

Let $\epsilon_P$ be a small counterclockwise circle around $P\in F$ in $T^2$,
and let $n_P[\alpha]:=\deg(\alpha|_{\epsilon_P})$ denote the winding number of $\alpha$.
For $\beta$, define the integer
\begin{align}
r_P[\beta]:=\frac12\deg(\beta|_{\epsilon_P})
=\frac1{4\pi}\oint_{\epsilon_P}d\beta
=\frac{n_P[\alpha']-n_P[\alpha]}2.
\label{eq:beta_half_vorticity}
\end{align}
The integral is taken for a piecewise smooth representative.
The winding number is even because $\beta_{-\bk}=\beta_\bk$, so $r_P[\beta]\in\Z$.
Since $\epsilon_P$ covers a small circle in the quotient space twice,
$r_P$ equals the ordinary winding number on $Q^\circ$. Hence
\begin{align}
\sum_{P\in F}r_P[\beta]=0.
\label{eq:beta_vorticity_sum}
\end{align}
Subject to this constraint, these four integers determine the homotopy class of $\beta$.

Using the partial polarization along the path $\G\to Y\to M$, write
$\nu^{C_2}_{1,\G\to M}:=\nu^{C_2}_{1,\G\to Y}+\nu^{C_2}_{1,Y\to M}\pmod2$.
The invariant then changes as
\begin{align}
\nu_2^{C_2}[\alpha+\beta]-\nu_2^{C_2}[\alpha]
&=r_X[\beta]\nu^{C_2}_{1,\G\to X}
 +r_Y[\beta]\nu^{C_2}_{1,\G\to Y}
 +r_M[\beta]\nu^{C_2}_{1,\G\to M}\pmod2.
\label{eq:alpha_general_change_explicit}
\end{align}
The proof is given in Appendix~\ref{app:alpha_change}.

\subsection{Atomic insulators and topological indices}
\label{sec:atomic_c2}

We calculate the indices of atomic insulators for $\alpha=\alpha^0$,
defined in \eqref{eq:canonical_alpha}, and then examine the change
to $\alpha=\alpha^1$, defined in \eqref{eq:alternative_alpha}.

As in Sec.~\ref{sec:c2t_model}, consider the four high-symmetry
positions in the unit cell,
\begin{align}
({\rm x},{\rm y})\in
\left\{(0,0),\left(\tfrac12,0\right),
\left(0,\tfrac12\right),\left(\tfrac12,\tfrac12\right)\right\}.
\end{align}
At each position, we place two occupied orbitals forming a Kramers pair.
We denote the resulting atomic insulator by
${\mathsf A}_{{\rm x}{\rm y}}$, distinguishing it from the one-band
model ${\mathsf a}_{{\rm x}{\rm y}}$ introduced earlier.
The partial polarization $\g_{\rm T}/(2\pi)$ gives the Wannier center
of one member of the pair in the corresponding direction. Hence
\begin{align}
\nu^{C_2}_{1,\G\to X}({\mathsf A}_{{\rm x}{\rm y}})
&=\nu^{C_2}_{1,Y\to M}({\mathsf A}_{{\rm x}{\rm y}})
=2{\rm x},\\
\nu^{C_2}_{1,\G\to Y}({\mathsf A}_{{\rm x}{\rm y}})
&=2{\rm y}
\qquad(\bmod\,2).
\end{align}
These indices distinguish the four atomic insulators.
However, the direct sums
${\mathsf A}_{00}\oplus{\mathsf A}_{\frac12\frac12}$ and
${\mathsf A}_{\frac120}\oplus{\mathsf A}_{0\frac12}$
have the same band number and the same one-dimensional indices.
We show below that $\nu_2^{C_2}$ distinguishes them.

For all four atomic insulators, choose
\begin{align}
D(C_2T)=1_2,\qquad D(C_2)=i\s_y.
\end{align}
The orbital position enters through the reciprocal-lattice
translation matrix,
\begin{align}
V_\bG=e^{-i(G_x{\rm x}+G_y{\rm y})}1_2.
\end{align}
At a high-symmetry point, the gauge condition
\eqref{eq:c2_gauge} reads
\begin{align}
V_{2P}D(C_2)\Phi^P_P=\Phi^P_PJ,
\qquad P\in\{\G,X,Y,M\}.
\end{align}
Thus we may choose a constant frame near $P$:
$\Phi^P_\bk=\pm1_2$ when $V_{2P}=1_2$, and
$\Phi^P_\bk=\pm\s_z$ when $V_{2P}=-1_2$.
Frames on translated patches are determined by
$\Phi^P_{\bk+\bG}=V_\bG\Phi^P_\bk$.

\begin{figure}[tb]
\centering
\includegraphics[width=0.8\linewidth]{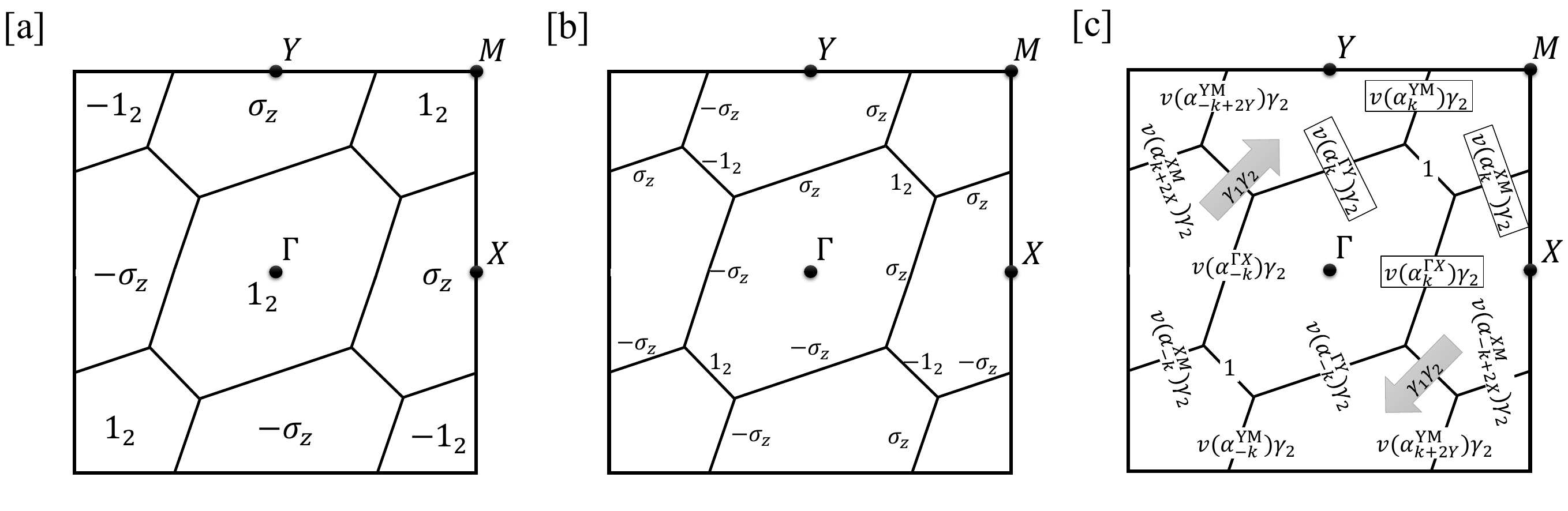}
\caption{Calculation of $\nu_2^{C_2}[\alpha]$ for
${\mathsf A}_{\frac12\frac12}$.
[a] Local Bloch frames $\Phi^P_\bk$.
[b] Transition functions $t^{PQ}_\bk$ on double overlaps.
[c] Continuous ${\rm Pin}_+(2)$ lifts after the singular gauge
transformation.
Boxed expressions specify the lifts on the four reference
components with reflection-type transition functions.
Here $\alpha^{PQ}$ is a real-valued branch chosen on the
reference component $U_{PQ}$.
On each symmetry-related component, the momentum argument is
mapped back to the reference component, ensuring $C_2$ symmetry
and periodicity of the lift.
Gray arrows indicate the direction of the transition $X\to Y$;
the reverse transition carries the Hermitian conjugate lift.}
\label{fig:c2patch_ai}
\end{figure}

Consider ${\mathsf A}_{\frac12\frac12}$.
Since $V_{2X}=V_{2Y}=-1_2$, its boundary conditions are
antiperiodic in both directions.
Figures~\ref{fig:c2patch_ai}[a] and [b] show a choice of local
frames and their transition functions.
Here $t^{PQ}_\bk=t^{QP}_\bk$, so panel [b] omits the patch order.

We now construct the lifts shown in Fig.~\ref{fig:c2patch_ai}[c].
Under the Clifford convention \eqref{eq:clifford_convention},
the reflection $-\s_z$ lifts to $\g_1$.
Define
\begin{align}
v(\theta):=e^{\frac{\theta}{2}\g_1\g_2},
\qquad v(\theta+2\pi)=-v(\theta).
\end{align}
For a real angle $\theta\equiv\alpha_\bk\pmod{2\pi}$,
the local lifting rules are
\begin{align}
t^{PQ}_\bk=1_2
&\longmapsto\tilde t^{PQ}_\bk=1_2
\longmapsto\pm1,\\
t^{PQ}_\bk=-1_2
&\longmapsto\tilde t^{PQ}_\bk=-1_2
\longmapsto\pm\g_1\g_2,\\
t^{PQ}_\bk=\s_z
&\longmapsto\tilde t^{PQ}_\bk=e^{i\alpha_\bk\s_y}\s_z
\longmapsto\pm v(\theta)\g_2,\\
t^{PQ}_\bk=-\s_z
&\longmapsto\tilde t^{PQ}_\bk=e^{i\alpha_\bk\s_y}(-\s_z)
\longmapsto\pm v(\theta)\g_1.
\end{align}
Since $(v(\theta)\g_a)^\dagger=v(\theta)\g_a$ for $a=1,2$,
only the lift $\g_1\g_2$ changes sign when the patch order is reversed.

To impose \eqref{eq:c2sym_lift_cond}, first choose a reference
component $U_{PQ}$ from each set of overlap components related
by $C_2$ rotations and reciprocal-lattice translations.
On this component, choose a continuous real-valued branch
\begin{align}
\alpha^{PQ}:U_{PQ}\to\R,\qquad
\alpha^{PQ}_\bk\equiv\alpha_\bk\pmod{2\pi}.
\end{align}
After choosing the lift there, define it on a symmetry-related
component by
\begin{align}
u(\tilde t^{PQ}_\bk)
:=u(\tilde t^{PQ}_{g^{-1}\bk}),
\qquad \bk\in g(U_{PQ}),
\end{align}
where $g$ is a $C_2$ rotation, a reciprocal-lattice translation,
or their composition.
Because $\tilde t^{PQ}$ is $C_2$ invariant and periodic, this
prescription gives a lift of the correct transition matrix.

For the four reflection-type overlaps, choose the reference
components boxed in Fig.~\ref{fig:c2patch_ai}[c], all of which
intersect $[0,\pi]\times[0,\pi]$.
Their transition functions are $\s_z$, and we choose
\begin{align}
u(\tilde t^{PQ}_\bk)=v(\alpha^{PQ}_\bk)\g_2,
\qquad PQ\in\{\G X,\G Y,XM,YM\},
\end{align}
on these reference components.
We also choose $u(\tilde t^{\G M})=1$ and
$u(\tilde t^{XY})=\g_1\g_2$ in the directions indicated by the arrows.

The four independent triple overlaps can be chosen in
$[-\pi,\pi]\times[0,\pi]$.
Using the cyclic patch order in each superscript, their lifted
products are
\begin{align}
(-1)^{z^{\G XY}_\alpha}
&=(v(\alpha^{\G X}_{-\bk})\g_2)
  (\g_1\g_2)(v(\alpha^{\G Y}_\bk)\g_2)
\notag\\
&=-v(\alpha^{\G X}_{-\bk}-\alpha^{\G Y}_\bk)\g_1\g_2,
\\
(-1)^{z^{\G XM}_\alpha}
&=(v(\alpha^{\G X}_\bk)\g_2)
  (v(\alpha^{XM}_\bk)\g_2)(1)
\notag\\
&=v(\alpha^{\G X}_\bk-\alpha^{XM}_\bk),
\\
(-1)^{z^{\G YM}_\alpha}
&=(v(\alpha^{\G Y}_\bk)\g_2)
  (v(\alpha^{YM}_\bk)\g_2)(1)
\notag\\
&=v(\alpha^{\G Y}_\bk-\alpha^{YM}_\bk),
\\
(-1)^{z^{XYM}_\alpha}
&=(\g_1\g_2)
  (v(\alpha^{YM}_{-\bk+2Y})\g_2)
  (v(\alpha^{XM}_{\bk+2X})\g_2)
\notag\\
&=v(\alpha^{YM}_{-\bk+2Y}-\alpha^{XM}_{\bk+2X})
  \g_1\g_2.
\label{eq:atomic_reference_triples}
\end{align}
Thus the calculation reduces to the four real-valued angle
differences appearing in these expressions.

For $\alpha=\alpha^0$, all four reference components admit
continuous branches satisfying
\begin{align}
-\pi<\alpha^{0,PQ}_\bk<\pi,
\qquad PQ\in\{\G X,\G Y,XM,YM\}.
\end{align}
On the corresponding triple overlaps, these branches give
\begin{align}
\alpha^{0,\G X}_{-\bk}-\alpha^{0,\G Y}_\bk&=-\pi,\\
\alpha^{0,\G X}_\bk-\alpha^{0,XM}_\bk&=0,\\
\alpha^{0,\G Y}_\bk-\alpha^{0,YM}_\bk&=0,\\
\alpha^{0,YM}_{-\bk+2Y}-\alpha^{0,XM}_{\bk+2X}&=-\pi.
\end{align}
Since $v(-\pi)=-\g_1\g_2$, the four lifted products in
\eqref{eq:atomic_reference_triples} are $(-1,1,1,1)$.
Only $U_{\G XY}$ therefore has a nonzero $\Z_2$ value, yielding
\begin{align}
\nu_2^{C_2}[\alpha^0]
({\mathsf A}_{\frac12\frac12})=1.
\end{align}

For $\alpha=\alpha^1$, the branches on the reference components
$\G X$, $\G Y$, and $XM$ can again be chosen in $(-\pi,\pi)$.
On the reference $YM$ component, however, the principal value
crosses its branch cut at $k_y=\pi$.
A continuous branch is instead obtained by choosing
\begin{align}
0<\alpha^{1,YM}_\bk<2\pi.
\end{align}
The first three angle differences remain $-\pi,0,0$, whereas
the last becomes
\begin{align}
\alpha^{1,YM}_{-\bk+2Y}
-\alpha^{1,XM}_{\bk+2X}=\pi.
\end{align}
The change from $-\pi$ to $\pi$ reverses the last lifted product,
because $v(\theta+2\pi)=-v(\theta)$.
The four products are now $(-1,1,1,-1)$, so both
$U_{\G XY}$ and $U_{XYM}$ have nonzero $\Z_2$ values. Hence
\begin{align}
\nu_2^{C_2}[\alpha^1]
({\mathsf A}_{\frac12\frac12})=0.
\end{align}

The general change formula
\eqref{eq:alpha_general_change_explicit} gives the same result.
For an atomic insulator and $\beta=\alpha-\alpha^0$,
\begin{align}
\nu_2^{C_2}[\alpha]({\mathsf A}_{{\rm x}{\rm y}})
&=4{\rm x}{\rm y}
+2{\rm x}\bigl(r_X[\beta]+r_M[\beta]\bigr)
\notag\\
&\quad
+2{\rm y}\bigl(r_Y[\beta]+r_M[\beta]\bigr)
\quad\bmod2.
\end{align}
For $\beta=\alpha^1-\alpha^0$, the winding numbers of
$\alpha^0$ and $\alpha^1$ at $(\G,X,Y,M)$ are
$(1,-1,-1,1)$ and $(1,-1,1,-1)$, respectively. Thus
\begin{align}
(r_\G,r_X,r_Y,r_M)=(0,0,1,-1).
\end{align}
The resulting indices are
\begin{align}
\begin{array}{c|cccc|cc}
&\nu_0
&\nu^{C_2}_{1,\G\to X}
&\nu^{C_2}_{1,Y\to M}
&\nu^{C_2}_{1,\G\to Y}
&\nu_2^{C_2}[\alpha^0]
&\nu_2^{C_2}[\alpha^1]\\ \hline
{\mathsf A}_{00}
&2&0&0&0&0&0\\
{\mathsf A}_{\frac120}
&2&1&1&0&0&1\\
{\mathsf A}_{0\frac12}
&2&0&0&1&0&0\\
{\mathsf A}_{\frac12\frac12}
&2&1&1&1&1&0
\end{array}
\label{eq:atomic_c2_indices}
\end{align}

Finally, additivity gives
\begin{align}
\nu_2^{C_2}[\alpha^0]
({\mathsf A}_{00}\oplus{\mathsf A}_{\frac12\frac12})&=1,\\
\nu_2^{C_2}[\alpha^0]
({\mathsf A}_{\frac120}\oplus{\mathsf A}_{0\frac12})&=0.
\end{align}
For $\alpha^1$, these values become $0$ and $1$, respectively.
Thus either choice distinguishes the two direct sums,
although their band numbers and one-dimensional indices coincide.

\subsection{Discrete formulation}
\label{sec:w2_c2}
We extend the discrete formula for the second SW number in
Sec.~\ref{sec:dis_w2}~\cite{ShiozakiDiscrete2024} to compute $\nu_2^{C_2}[\alpha]$
for the function $\alpha_\bk$ chosen in Sec.~\ref{sec:c2z2top_torus}.
The input consists of occupied Bloch frames chosen independently at discrete
mesh points; no smooth gauge between mesh points is needed.
The dependence on $\alpha_\bk$ enters through the rules relating lifts on paired
edges of the half-BZ boundary.

\subsubsection{Gauge-fixing conditions}
We first rewrite the continuum lift conditions in terms of the transition
functions before the singular gauge transformation.
Let ${\cal D}:=[-\pi,\pi]\times[0,\pi]$ be the half BZ, and let
${\cal D}^{\circ}$ be obtained from ${\cal D}$ by removing the high-symmetry points and their
reciprocal-lattice images, namely $(0,0),(0,\pi),(\pm\pi,0),(\pm\pi,\pi)$.
The removed points lie on the boundary, and ${\cal D}^{\circ}$ is contractible.
We can therefore choose a continuous real-valued branch of $\alpha_\bk$,
\begin{align}
\tilde\alpha:{\cal D}^{\circ}\longrightarrow\R,
\qquad \tilde\alpha_\bk\equiv\alpha_\bk\pmod{2\pi}.
\label{eq:dis_alpha_real_branch}
\end{align}
All branch differences below are evaluated using this single branch.

For the standard antisymmetric matrix $J = \bigoplus_{a=1}^m (i\sigma_y)$, define the continuous Spin lift of the rotation $e^{-\theta J/2}$ by
\begin{align}
h(\theta):=u(e^{-\theta J/2}),\qquad
\theta\in\R,\quad h(0)=1.
\end{align}
Here $\theta$ is treated as a real number.
This lift satisfies $h(\theta)h(\theta')=h(\theta+\theta')$ and
$h(\theta)^\dagger=h(-\theta)$. In particular,
\begin{align}
h(-\pi)=u(J),\qquad h(\pi)=h(-\pi)^\dagger.
\label{eq:dis_v_convention}
\end{align}

Using local frames satisfying \eqref{eq:c2_gauge}, we may choose the lift of
\eqref{eq:transition_func_singular_gauge} as
\begin{align}
u(\tilde t^{PQ}_\bk)
=h(\tilde\alpha_\bk)^\dagger
u(t^{PQ}_\bk)h(\tilde\alpha_\bk).
\label{eq:t_lift_ba}
\end{align}
The intermediate factors of $h$ cancel in the product on a triple overlap,
so its $\Z_2$ sign is unchanged by the transformation.
For two boundary points $\bk,\bk'$ identified within ${\cal D}^{\circ}$ with $C_2$ rotation or reciprocal lattice translation, the lift condition \eqref{eq:c2sym_lift_cond} is equivalent to
\begin{align}
u(t^{PQ}_{\bk'})
=h(\tilde\alpha_{\bk'}-\tilde\alpha_\bk)
u(t^{PQ}_\bk)
h(\tilde\alpha_{\bk'}-\tilde\alpha_\bk)^\dagger.
\label{eq:dis_transition_boundary_rule}
\end{align}
On the upper boundary, we use $\bk'= -\bk+(0,2\pi)$.

Specifically, define the boundary branch differences by
\begin{align}
\Delta_j&:=\tilde\alpha_{(-x,j)}-\tilde\alpha_{(x,j)}
\in\pi+2\pi\Z,
&&0<x<\pi,\quad j=0,\pi,
\label{eq:dis_alpha_jump_horizontal}\\
\Delta_x&:=\tilde\alpha_{(-\pi,y)}-\tilde\alpha_{(\pi,y)}
\in2\pi\Z,
&&0<y<\pi.
\label{eq:dis_alpha_jump_vertical}
\end{align}
These are continuous functions with discrete values and are therefore constant
on each open interval.
Thus all information about $\alpha_\bk$ needed for the calculation is contained
in the three constants $\Delta_0,\Delta_\pi,\Delta_x$.
Adding an integer multiple of $2\pi$ to the entire branch leaves them unchanged.

For the reference function \eqref{eq:canonical_alpha}, choosing
$0\leq\tilde\alpha^0\leq\pi$ on ${\cal D}^{\circ}$ gives
\begin{align}
\Delta_0=\Delta_\pi=\pi,\qquad \Delta_x=0.
\label{eq:dis_reference_jumps}
\end{align}

\subsubsection[Discrete definition of the Z2 invariant]{Discrete definition of the $\Z_2$ invariant}
\begin{figure}[tb]
\centering
\includegraphics[width=\linewidth]{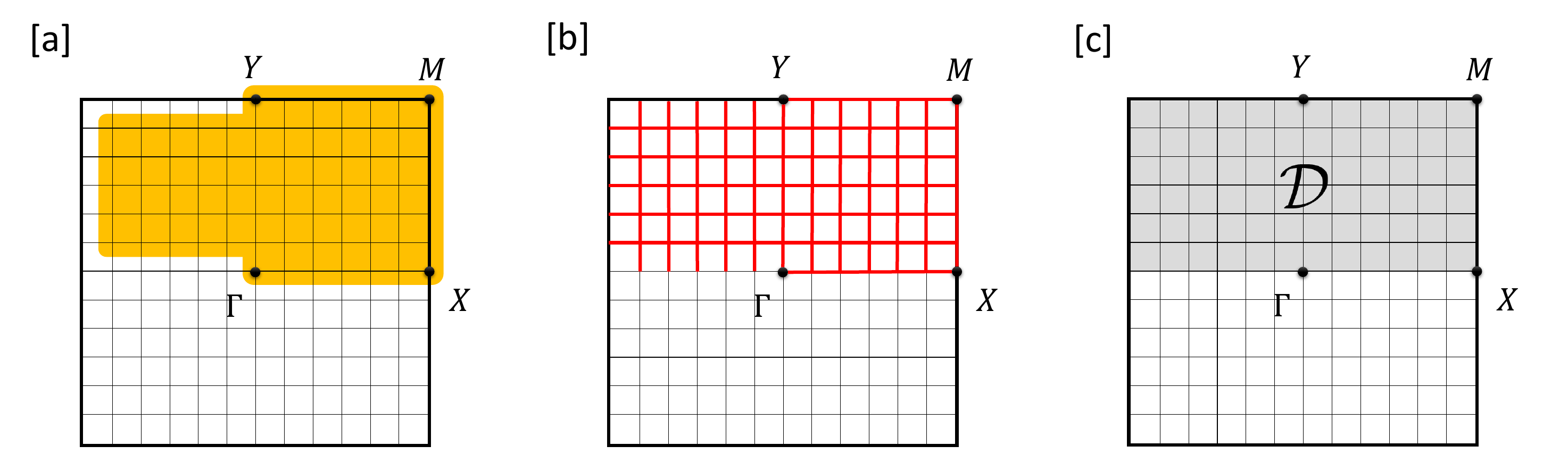}
\caption{Vertices, edges, and plaquettes used to compute $\nu_2^{C_2}[\alpha]$
on a discrete mesh, illustrated for $L=6$.
[a] Vertices ${\cal V}$ at which frames are chosen independently, shown as mesh points in the yellow region.
Black dots indicate the high-symmetry points $\G,X,Y,M$.
[b] Edges ${\cal E}$ on which lifts are chosen independently, shown in red.
[c] Half-BZ domain ${\cal D}$ over which the plaquette values are summed.
The remaining boundary frames are fixed by symmetry, and the edge lifts are
determined by the branch differences $\Delta_0,\Delta_\pi,\Delta_x$ of the chosen
function $\alpha_\bk$.}
\label{fig:c2_ref_pts}
\end{figure}

Divide ${\cal D}$ into $2L\times L$ plaquettes and write
\begin{align}
\bk_{\bm n}:=\frac{\pi}{L}(n_x,n_y),\qquad
\Phi_{\bm n}:=\Phi_{\bk_{\bm n}},\qquad
\bm e_x=(1,0),\quad\bm e_y=(0,1).
\end{align}
Choose the set of independent vertices as
\begin{align}
{\cal V}
&:=\{\bk_{\bm n}\mid -L+1\leq n_x\leq L,\ 1\leq n_y\leq L-1\}
\notag\\
&\quad\cup\{\bk_{\bm n}\mid 0\leq n_x\leq L,\ n_y\in\{0,L\}\}.
\label{eq:dis_vertex_set}
\end{align}
The high-symmetry points in this set are ${\cal V}_{\rm HSP}=\{\G,X,Y,M\}$.
The set ${\cal E}$ of independent oriented edges consists of the following
three types:
\begin{align}
\begin{array}{c|c|c}
\text{Edge}&n_x&n_y\\ \hline
(\bk_{\bm n},\bk_{\bm n+\bm e_x})&-L,\ldots,L-1&1,\ldots,L-1\\
(\bk_{\bm n},\bk_{\bm n+\bm e_y})&-L+1,\ldots,L&0,\ldots,L-1\\
(\bk_{\bm n},\bk_{\bm n+\bm e_x})&0,\ldots,L-1&0,L
\end{array}
\label{eq:dis_edge_set}
\end{align}
Thus we choose all interior edges, the right halves of the upper and lower
boundaries, and the right boundary, as shown in Fig.~\ref{fig:c2_ref_pts}.

At each independent vertex, take an arbitrary orthonormal occupied frame as input
and transform it to the real gauge by a change of basis at that vertex\footnote{For an input frame,
$w_\bk:=\Phi_\bk^\dagger D(C_2T)\Phi_\bk^*$ is a symmetric unitary matrix.
A Takagi factorization $w_\bk=R_\bk R_\bk^\top$ gives a frame
$\Phi_\bk R_\bk$ satisfying the real-gauge condition.
This operation can be performed independently at each vertex.}:
\begin{align}
D(C_2T)(\Phi_{\bm n})^*=\Phi_{\bm n}.
\label{eq:dis_c2_gauge_1}
\end{align}
At the high-symmetry points, also impose
\begin{align}
V_{2P}D(C_2)\Phi_P=\Phi_PJ,
\qquad P\in\{\G,X,Y,M\}.
\label{eq:dis_c2_gauge_cond}
\end{align}
Indeed, in the real gauge, $\Phi_P^\dagger V_{2P}D(C_2)\Phi_P$ is a real
orthogonal matrix whose square is $-1$ and can therefore be brought to $J$
by an orthogonal change of basis.
These are pointwise gauge choices and require no continuity between neighboring
vertices.

Define the frames at the remaining boundary vertices by
\begin{align}
\Phi_{(-n,0)}&=D(C_2)\Phi_{(n,0)}J^\top,
&&1\leq n\leq L-1,\notag\\
\Phi_{(-n,L)}&=V_{(0,2\pi)}D(C_2)\Phi_{(n,L)}J^\top,
&&1\leq n\leq L-1,\notag\\
\Phi_{(-L,n_y)}&=V_{(-2\pi,0)}\Phi_{(L,n_y)},
&&0\leq n_y\leq L.
\label{eq:dis_c2_gauge_2}
\end{align}
Consistency at the high-symmetry points follows from \eqref{eq:dis_c2_gauge_cond}.

For each edge, take the orthogonal factor of the overlap matrix obtained by
singular value decomposition:
\begin{align}
\Phi_{\bm n}^\dagger\Phi_{\bm n+\bm e_\mu}
=L\Sigma R^\top,
\qquad t^\mu_{\bm n}:=L R^\top\in O(2m).
\label{eq:dis_orthogonal_link}
\end{align}
We assume that every overlap matrix is nonsingular, so its orthogonal factor
is unique. At the boundary, these factors satisfy
\begin{align}
t^x_{(-n-1,j)}&=J(t^x_{(n,j)})^\top J^\top,
&&0\leq n\leq L-1,\quad j=0,L,
\label{eq:dis_t_c2}\\
t^y_{(-L,n_y)}&=t^y_{(L,n_y)},
&&0\leq n_y\leq L-1.
\label{eq:dis_t_rlv}
\end{align}

Choose a ${\rm Pin}_+(2m)$ lift on each independent edge and write
$U^\mu_{\bm n}:=u(t^\mu_{\bm n})$.
Following \eqref{eq:dis_transition_boundary_rule}, define the lifts on the
remaining boundary edges by
\begin{align}
U^x_{(-n-1,j)}
&:=h(\Delta_{\pi j/L})
(U^x_{(n,j)})^\dagger h(\Delta_{\pi j/L})^\dagger,
&&j=0,L,
\label{eq:dis_c2_lift_general}\\
U^y_{(-L,n_y)}
&:=h(\Delta_x)U^y_{(L,n_y)}h(\Delta_x)^\dagger.
\label{eq:dis_periodic_lift_general}
\end{align}
The index ranges are the same as in \eqref{eq:dis_t_c2} and \eqref{eq:dis_t_rlv}.
The Hermitian conjugate on horizontal edges appears because $C_2$ reverses their
orientation. A reversed edge always carries the Hermitian conjugate lift.

The orthogonal matrices covered by $h(\Delta_0),h(\Delta_\pi)$ are $\pm J$,
whereas that covered by $h(\Delta_x)$ is $\pm1_{2m}$.
The lift rules are therefore compatible with \eqref{eq:dis_t_c2} and
\eqref{eq:dis_t_rlv}.
Since each $\Delta$ is a constant defined on an open interval, the same rule also
applies to edges incident on a high-symmetry point.

For $\alpha=\alpha^0$, these rules reduce to
\begin{align}
U^x_{(-n-1,j)}=h(-\pi)^\dagger(U^x_{(n,j)})^\dagger h(-\pi),
\qquad
U^y_{(-L,n_y)}=U^y_{(L,n_y)}.
\label{eq:dis_c2_lift_corrected}
\end{align}
The conjugation takes the form $h(-\pi)^\dagger(\cdot)h(-\pi)$ because $\Delta_0=\Delta_\pi=\pi$.

Form the product around each plaquette:
\begin{align}
{\cal U}_{\Box_{\bm n}}
:=U^x_{\bm n}U^y_{\bm n+\bm e_x}
(U^x_{\bm n+\bm e_y})^\dagger(U^y_{\bm n})^\dagger.
\end{align}
Whether this product is closer to $1$ or $-1$ is determined by
\begin{align}
(-1)^{z_{\Box_{\bm n}}}
:=\operatorname{sgn}\operatorname{Re}\Tr{\cal U}_{\Box_{\bm n}}.
\label{eq:dis_c2_sign}
\end{align}
We use a unitary representation of the Clifford algebra and assume that the
right-hand side is nonzero. The discrete invariant is
\begin{align}
\nu_2^{C_2,{\rm dis}}[\alpha]
:=\sum_{n_x=-L}^{L-1}\sum_{n_y=0}^{L-1}
z_{\Box_{\bm n}}\quad\bmod2.
\label{eq:dis_c2_invariant}
\end{align}
In numerical calculation, the smallest overlap singular value should be recorded to check
nonsingularity.

For a sufficiently fine mesh,
\begin{align}
\nu_2^{C_2,{\rm dis}}[\alpha]=\nu_2^{C_2}[\alpha].
\label{eq:dis_c2_continuum_match}
\end{align}
To establish this equality, first consider plaquettes away from
the fixed points.
Writing $h_a:=h(\tilde\alpha_{\bk_a})$, the singular gauge
transformation acts on edge lifts as
$U_{ab}\mapsto\tilde U_{ab}=h_a^\dagger U_{ab}h_b$.
The intermediate factors cancel,
$\tilde U_{ab}\tilde U_{bc}=h_a^\dagger U_{ab}U_{bc}h_c$,
so each plaquette product changes only by conjugation,
leaving $z_\Box$ unchanged.
Near each fixed point, a symmetry-preserving local deformation
makes the occupied projector constant without closing the band gap.
We may then choose a constant frame and edge lifts equal to $1$,
so every plaquette contained in this constant-projector region
has $z_\Box=0$.
In the continuous construction, these neighborhoods are trivial
local patches, glued to neighboring patches by $\tilde t^{PQ}$
on overlaps that avoid the fixed points.
Thus no value of $h$ at a fixed point is needed.
Together with the boundary conditions
\eqref{eq:dis_c2_lift_general} and
\eqref{eq:dis_periodic_lift_general}, this identifies the
plaquette-sign sum with the ordinary discrete second SW
number of the real bundle on $T^2/C_2$.
The mesh is chosen sufficiently fine throughout the local
deformation so that no plaquette trace passes through zero.

\subsubsection{Gauge invariance}
Consider an allowed frame change $\Phi_{\bm n}\mapsto\Phi_{\bm n}W_{\bm n}$,
with $W_{\bm n}\in O(2m)$ and $[W_P,J]=0$ at the high-symmetry points.
The group of real orthogonal matrices commuting with $J$ is isomorphic to the
connected group $U(m)$.
Thus the lift of $W_P$ commutes with $u(J)$ and also with
$h(\Delta_j),h(\Delta_x)$.
To see this, lift a path from $1$ to $W_P$ in $U(m)$. 
Its commutator with $v$ is a continuous $\{\pm1\}$-valued function, equal to $+1$
at the starting point and hence also $+1$ at the endpoint.

Choose $w_{\bm n}:=u(W_{\bm n})$ at the independent vertices and define it at
the remaining vertices by
\begin{align}
w_{(-n,j)}
&:=h(\Delta_{\pi j/L})w_{(n,j)}h(\Delta_{\pi j/L})^\dagger,
&&1\leq n\leq L-1,\quad j=0,L,\notag\\
w_{(-L,n_y)}
&:=h(\Delta_x)w_{(L,n_y)}h(\Delta_x)^\dagger,
&&0\leq n_y\leq L.
\end{align}
Commutativity at the high-symmetry points ensures consistency at the boundary
endpoints.

On any oriented edge $(a,b)$, the transformed lift can be written as
\begin{align}
U_{ab}\longmapsto(-1)^{\eta_{ab}}w_a^\dagger U_{ab}w_b,
\qquad\eta_{ab}\in\Z_2.
\end{align}
The boundary rules assign the same $\eta$ to two identified edges.
In a plaquette product, the factors of $w$ at intermediate vertices cancel,
and conjugation at the base point leaves the trace unchanged. Therefore,
\begin{align}
z_\Box\longmapsto z_\Box+\sum_{e\subset\partial\Box}\eta_e
\quad\bmod2.
\end{align}
In the sum over ${\cal D}$, each interior-edge contribution $\eta$ appears twice,
and each boundary-edge contribution $\eta$ cancels with that of its identified partner.
Thus \eqref{eq:dis_c2_invariant} is independent of the frame choices.
Setting $W_{\bm n}=1$ also proves independence of the signs chosen for the lifts.

\subsubsection{Numerical examples}
\label{sec:c2_numerical_examples}

We apply the discrete formula \eqref{eq:dis_c2_invariant} to atomic insulators
and time-reversal-symmetric topological insulators.
We use the reference function $\alpha^0$ and fix the boundary lifts by
\eqref{eq:dis_c2_lift_corrected}.

For the atomic insulators ${\mathsf A}_{{\rm x}{\rm y}}$, take $H_\bk=-1_2$
with the symmetry and reciprocal-lattice translation matrices of
Sec.~\ref{sec:atomic_c2}.
For a TI example, consider the four-band model
\begin{align}
H_\bk(\mu)
&=\sin k_x\,\s_z\otimes\s_0
+\sin k_y\,\s_x\otimes\s_0\notag\\
&\quad +(\mu-\cos k_x-\cos k_y)\,\s_y\otimes\s_y.
\label{eq:numerical_ti_model}
\end{align}
Here $\s_0=1_2$, $\mu$ is a mass parameter, and
\begin{align}
D(C_2T)=1_4,\qquad
D(C_2)=i\s_y\otimes\s_0,\qquad V_\bG=1_4.
\end{align}
This Hamiltonian is real symmetric and satisfies the required symmetries.
Its eigenvalues, each doubly degenerate, are
\begin{align}
E_\pm(\bk)
=\pm\sqrt{\sin^2k_x+\sin^2k_y
+(\mu-\cos k_x-\cos k_y)^2}.
\end{align}
We occupy the two negative-energy bands and use the gapped parameter values
$\mu=1,-1,3$.

The TI character can also be checked analytically.
In the $+1$ eigensector of the conserved operator $\s_0\otimes\s_y$, the Hamiltonian is
$\bm d\cdot\bm\s$, with $\bm d=(\sin k_y,\mu-\cos k_x-\cos k_y,\sin k_x)$.
The Dirac masses and orientations at the four high-symmetry points give the
Chern number of the negative-energy band as
\begin{align}
C_+=\frac12\bigl[\sgn(\mu-2)+\sgn(\mu+2)-2\sgn\mu\bigr],
\qquad\mu\notin\{-2,0,2\}.
\label{eq:ti_sector_chern}
\end{align}
We use the Berry connection and BZ orientation specified in \eqref{eq:berry_convention}.
Since time reversal exchanges the two sectors, $\nu_{\rm KM}=C_+\pmod2$
~\cite{Kane22005}.
Thus $\mu=\pm1$ gives TI phases, whereas $\mu=3$ gives a trivial phase.

We diagonalize the Hamiltonian independently at each mesh point and impose
\eqref{eq:dis_c2_gauge_cond} on the resulting real occupied frames at the
high-symmetry points.
No gauge smoothing is performed between neighboring points.
We obtain the orthogonal links by singular value decomposition of the overlap
matrices and construct their ${\rm Pin}_+$ lifts by decomposing them into
Householder reflections.
The one-dimensional invariants are calculated from the same links, and the
Kane--Mele index is obtained from \eqref{eq:km_partial_indices}.

Table~\ref{tab:c2_numerical_examples} lists the results for a half BZ divided
into $2L\times L$ plaquettes.
All indices in the table agree for $L=8,16,32$.
The atomic-insulator results reproduce the analytical values in
\eqref{eq:atomic_c2_indices}.
The models $H(1)$ and $H(-1)$ have $\nu_{\rm KM}=1$ and are TI phases,
but have $\nu_2^{C_2,{\rm dis}}[\alpha^0]=0$.
All $\Z_2$ indices vanish for $H(3)$.
The direct sum of $H(1)$ and ${\mathsf A}_{\frac12\frac12}$ has
$\nu_2^{C_2,{\rm dis}}[\alpha^0]=1$ while retaining $\nu_{\rm KM}=1$.

\begin{table}[tb]
\centering
\caption{Invariants of atomic insulators and TI models computed from the
discrete formula. Here $H(\mu)$ denotes the occupied bundle of the negative-energy
bands of \eqref{eq:numerical_ti_model}.
The final column gives $\nu_2^{C_2,{\rm dis}}[\alpha^0]$.
All results agree for $L=8,16,32$.}
\label{tab:c2_numerical_examples}
\begin{tabular}{c|cccccc}
&$\nu_0$&$\nu^{C_2}_{1,\G\to X}$&$\nu^{C_2}_{1,Y\to M}$
&$\nu^{C_2}_{1,\G\to Y}$&$\nu_{\rm KM}$
&$\nu_2^{C_2,{\rm dis}}[\alpha^0]$\\ \hline
${\mathsf A}_{00}$&2&0&0&0&0&0\\
${\mathsf A}_{\frac120}$&2&1&1&0&0&0\\
${\mathsf A}_{0\frac12}$&2&0&0&1&0&0\\
${\mathsf A}_{\frac12\frac12}$&2&1&1&1&0&1\\ \hline
$H(1)$&2&1&0&1&1&0\\
$H(-1)$&2&0&1&0&1&0\\
$H(3)$&2&0&0&0&0&0\\
$H(1)\oplus{\mathsf A}_{\frac12\frac12}$&4&0&1&0&1&1
\end{tabular}
\end{table}

We also directly calculate the two atomic direct sums and obtain
\begin{align}
\nu_2^{C_2,{\rm dis}}[\alpha^0]
({\mathsf A}_{00}\oplus{\mathsf A}_{\frac12\frac12})&=1,\\
\nu_2^{C_2,{\rm dis}}[\alpha^0]
({\mathsf A}_{\frac120}\oplus{\mathsf A}_{0\frac12})&=0.
\end{align}
These results also agree for $L=8,16,32$ and are consistent with the previous section.

These examples also show that the indices distinguish the known stable classes.
The rank-zero atomic differences
\begin{align}
&[\mathsf A_{\frac120}]-[\mathsf A_{00}],\notag\\
&[\mathsf A_{0\frac12}]-[\mathsf A_{00}],\notag\\
&[\mathsf A_{\frac12\frac12}]-[\mathsf A_{\frac120}]
 -[\mathsf A_{0\frac12}]+[\mathsf A_{00}].
\end{align}
have indices $(1,0,0,0),(0,1,0,0),(0,0,0,1)$, respectively, in the order
$(\nu^{C_2}_{1,\G\to X},\nu^{C_2}_{1,\G\to Y},\nu_{\rm KM},\nu_2^{C_2}[\alpha^0])$.
Moreover, the analytical result $\nu_{\rm KM}=1$ for
$[H(1)]-[\mathsf A_{00}]$ shows that it is independent of these three elements.
Since the known rank-zero $K$ group is $\Z_2^{\oplus4}$, these four elements
generate the entire group.
Together with the integer component $\nu_0/2$, they give the complete stable
classification.
Furthermore, \eqref{eq:alpha_general_change_explicit} only adds a linear combination
of the one-dimensional indices to $\nu_2^{C_2}$, so fixing a different
$\alpha$ preserves the completeness of the classification.

\section{Conclusion}
\label{sec:conclusion}

We constructed the additional index of two-dimensional spinful systems
with $C_2$ and $T$ symmetries as the second SW number
$\nu_2^{C_2}[\alpha]$ of a real bundle over $T^2/C_2$, specified by an auxiliary
function $\alpha$. We proved gauge invariance, additivity under direct sums,
and the transformation law under changes of $\alpha$.
Together with the three one-dimensional indices and the number of occupied
Kramers pairs, this index distinguishes the known stable classes.
In particular, it distinguishes direct sums of atomic insulators that have
the same one-dimensional indices and band number.

The discrete formula encodes the auxiliary function in the lifts on the
boundary of a half BZ. It can be evaluated from frames chosen independently
at each vertex and agrees with the continuum definition on sufficiently
fine meshes. In numerical examples of atomic and TI models, we confirmed
that the index is unchanged by mesh refinement, gauge changes, and changes
of lift signs.

\section*{Acknowledgments}
We thank Seishiro Ono for useful discussions.
We acknowledge the use of ChatGPT (GPT-6) for assistance with
theoretical discussions and editing the English text.
This work was supported by JSPS KAKENHI Grant Nos.~JP22H05118,
JP26H01305, and JP26K00629.

\appendix

\section[Dependence on the singular gauge transformation]{Dependence of $\nu_2^{C_2}$ on the singular gauge transformation}
\label{app:alpha_change}

We prove the transformation law~\eqref{eq:alpha_general_change_explicit}.
Changing the singular gauge transformation from $\alpha$ to $\alpha+\beta$
gives
\begin{align}
\tilde t^{PQ}_{\alpha+\beta,\bk}
&=e^{-\frac{\beta_\bk}{2}(i1_{2m}-J)}
  \tilde t^{PQ}_{\alpha,\bk}e^{\frac{\beta_\bk}{2}(i1_{2m}-J)}\notag\\
&=e^{\beta_\bk J/2}
  \tilde t^{PQ}_{\alpha,\bk}e^{-\beta_\bk J/2}.
\label{eq:beta_transition_change}
\end{align}
The conjugating matrix in the first line is single-valued on $Q^\circ$
but is generally not real. The second line uses a real orthogonal matrix,
whose sign changes when the branch of $\beta_\bk$ is shifted by $2\pi$.
Thus the change is generally not a single-valued orthogonal gauge
transformation and can change the invariant.

On each contractible connected component of a double overlap, independently choose a continuous real branch $\beta^{PQ}_\bk \in \R$ satisfying 
\begin{align}
\beta^{PQ}_\bk\equiv\beta_\bk\pmod{2\pi}.
\end{align}
Use the same branch on components related by $C_2$ or reciprocal-lattice
translations. We suppress $\bk$ below and denote a continuous Spin lift by
\begin{align}
v^{PQ}_\beta:=u(e^{\beta^{PQ}J/2}).
\end{align}
The lifts of the new transition functions can then be chosen as
\begin{align}
u(\tilde t^{PQ}_{\alpha+\beta})
=v^{PQ}_\beta u(\tilde t^{PQ}_\alpha)(v^{PQ}_\beta)^\dagger.
\label{eq:beta_transition_lift}
\end{align}
Under a branch change $\beta^{PQ}\mapsto\beta^{PQ}+2\pi n$, $n\in\Z$, we have
\begin{align}
u(e^{(\beta^{PQ}+2\pi n)J/2})
=\pm\Omega^n v^{PQ}_\beta,
\qquad n\in\Z.
\end{align}
Here $\Omega$ is the lift in Eq.~\eqref{eq:clifford_volume}, and the sign
under conjugation is given by Eq.~\eqref{eq:clifford_parity}.
The determinant parity $p^{PQ}$ is independent of $\alpha$.

On each connected component of a triple overlap $U_{PQR}:=U_P\cap U_Q\cap U_R$, write the branch differences as
\begin{align}
\beta^{QR}-\beta^{PQ}=2\pi n^{PQR}_{QR},\qquad
\beta^{PR}-\beta^{PQ}=2\pi n^{PQR}_{PR}.
\end{align}
The integers $n^{PQR}_{QR},n^{PQR}_{PR}$ are constant on this component.
Since the overall signs of Spin lifts cancel under conjugation,
\begin{align}
&u(\tilde t^{PQ}_{\alpha+\beta})u(\tilde t^{QR}_{\alpha+\beta})\notag\\
&=v^{PQ}_\beta u(\tilde t^{PQ}_\alpha)(v^{PQ}_\beta)^\dagger
  v^{QR}_\beta u(\tilde t^{QR}_\alpha)(v^{QR}_\beta)^\dagger\notag\\
&=v^{PQ}_\beta u(\tilde t^{PQ}_\alpha)
  \Omega^{n^{PQR}_{QR}}u(\tilde t^{QR}_\alpha)
  (v^{PQ}_\beta)^\dagger(\Omega^\dagger)^{n^{PQR}_{QR}}\notag\\
&=(-1)^{n^{PQR}_{QR}p^{PQ}}\Omega^{n^{PQR}_{QR}}
  v^{PQ}_\beta u(\tilde t^{PQ}_\alpha)u(\tilde t^{QR}_\alpha)
  (v^{PQ}_\beta)^\dagger(\Omega^\dagger)^{n^{PQR}_{QR}}\notag\\
&=(-1)^{n^{PQR}_{QR}p^{PQ}+z^{PQR}_\alpha}
  \Omega^{n^{PQR}_{QR}}v^{PQ}_\beta
  u(\tilde t^{PR}_\alpha)(v^{PQ}_\beta)^\dagger
  (\Omega^\dagger)^{n^{PQR}_{QR}}\notag\\
&=(-1)^{n^{PQR}_{QR}p^{PQ}+z^{PQR}_\alpha}
  \Omega^{n^{PQR}_{QR}-n^{PQR}_{PR}}
  u(\tilde t^{PR}_{\alpha+\beta})
  (\Omega^\dagger)^{n^{PQR}_{QR}-n^{PQR}_{PR}}\notag\\
&=(-1)^{z^{PQR}_\alpha+n^{PQR}_{QR}p^{PQ}
 +(n^{PQR}_{QR}-n^{PQR}_{PR})p^{PR}}
  u(\tilde t^{PR}_{\alpha+\beta}).
\end{align}
Therefore,
\begin{align}
z^{PQR}_{\alpha+\beta}-z^{PQR}_\alpha
&=n^{PQR}_{QR}p^{PQ}
 +(n^{PQR}_{QR}-n^{PQR}_{PR})p^{PR}\notag\\
&=n^{PQR}_{QR}p^{QR}+n^{PQR}_{PR}p^{PR}
\quad\bmod2.
\label{eq:beta_cech_change}
\end{align}

Summing over the four independent triple overlaps gives
\begin{align}
\nu_2^{C_2}[\alpha+\beta]-\nu_2^{C_2}[\alpha]
&=n^{\G XY}_{XY}p^{XY}+n^{\G XY}_{\G Y}p^{\G Y}\notag\\
&\quad+n^{\G XM}_{XM}p^{XM}+n^{\G XM}_{\G M}p^{\G M}\notag\\
&\quad+n^{\G YM}_{YM}p^{YM}+n^{\G YM}_{\G M}p^{\G M}\notag\\
&\quad+n^{XYM}_{YM}p^{YM}+n^{XYM}_{XM}p^{XM}
\quad\bmod2.
\end{align}
Substituting
\begin{align}
p^{XY}&=p^{\G X}+p^{\G Y},\notag\\
p^{\G M}&=p^{\G Y}+p^{YM},\notag\\
p^{XM}&=p^{\G X}+p^{\G Y}+p^{YM}
\quad\bmod2
\end{align}
and using the freedom to reverse signs modulo two, we rewrite the coefficients as sums associated with closed paths:
\begin{align}
\nu_2^{C_2}[\alpha+\beta]-\nu_2^{C_2}[\alpha]
&\equiv
\bigl(n^{\G XY}_{XY}-n^{\G XM}_{XM}+n^{XYM}_{XM}\bigr)p^{\G X}
\notag\\
&\quad+\bigl(n^{\G XY}_{XY}-n^{\G XY}_{\G Y}
-n^{\G XM}_{XM}+n^{\G XM}_{\G M}
-n^{\G YM}_{\G M}+n^{XYM}_{XM}\bigr)p^{\G Y}
\notag\\
&\quad+\bigl(n^{\G XM}_{XM}-n^{\G XM}_{\G M}
-n^{\G YM}_{YM}+n^{\G YM}_{\G M}
+n^{XYM}_{YM}-n^{XYM}_{XM}\bigr)p^{YM}
\pmod2.
\end{align}

We now express each coefficient as a winding number. Fix a point in each
triple overlap and let $\beta^{PQ}|_{U_{PQR}}$ denote the value at that point.
For two points in the same double overlap, the difference between the values
of the chosen branch $\beta^{PQ}$ is the angular change along a path within
that overlap. Summing these changes around a closed path gives $2\pi$ times
the winding number of $\beta$ along that path in $Q^\circ$.
Reversing the path changes only the sign, which is immaterial
modulo two.

To be explicit, for the first coefficient, a closed path around $X$ gives
\begin{align}
&n^{\G XY}_{XY}-n^{\G XM}_{XM}+n^{XYM}_{XM}\notag\\
&=\frac1{2\pi}\Bigl[
 (\beta^{XY}-\beta^{\G X})|_{U_{\G XY}}
 -(\beta^{XM}-\beta^{\G X})|_{U_{\G XM}}
 +(\beta^{XM}-\beta^{XY})|_{U_{XYM}}\Bigr]\notag\\
&=\frac1{2\pi}\Bigl[
 (\beta^{XY}|_{U_{\G XY}}-\beta^{XY}|_{U_{XYM}})
 +(\beta^{\G X}|_{U_{\G XM}}-\beta^{\G X}|_{U_{\G XY}})\notag\\
&\hspace{35mm}
 +(\beta^{XM}|_{U_{XYM}}-\beta^{XM}|_{U_{\G XM}})\Bigr]\notag\\
&\equiv r_X[\beta]\pmod2.
\end{align}
The three bracketed differences are the angular changes along the
$XY$, $\G X$, and $XM$ double overlaps, respectively.
Similarly, a closed path enclosing both $Y$ and $M$ gives the second coefficient:
\begin{align}
&n^{\G XY}_{XY}-n^{\G XY}_{\G Y}
-n^{\G XM}_{XM}+n^{\G XM}_{\G M}
-n^{\G YM}_{\G M}+n^{XYM}_{XM}\notag\\
&=\frac1{2\pi}\Bigl[
 (\beta^{XY}-\beta^{\G Y})|_{U_{\G XY}}
 +(\beta^{\G M}-\beta^{XM})|_{U_{\G XM}}\notag\\
&\hspace{35mm}
 +(\beta^{\G Y}-\beta^{\G M})|_{U_{\G YM}}
 +(\beta^{XM}-\beta^{XY})|_{U_{XYM}}\Bigr]\notag\\
&=\frac1{2\pi}\Bigl[
 (\beta^{XY}|_{U_{\G XY}}-\beta^{XY}|_{U_{XYM}})
 +(\beta^{\G Y}|_{U_{\G YM}}-\beta^{\G Y}|_{U_{\G XY}})\notag\\
&\hspace{35mm}
 +(\beta^{\G M}|_{U_{\G XM}}-\beta^{\G M}|_{U_{\G YM}})
 +(\beta^{XM}|_{U_{XYM}}-\beta^{XM}|_{U_{\G XM}})\Bigr]\notag\\
&\equiv r_Y[\beta]+r_M[\beta]\pmod2.
\end{align}
For the third coefficient, a closed path around $M$ gives
\begin{align}
&n^{\G XM}_{XM}-n^{\G XM}_{\G M}
-n^{\G YM}_{YM}+n^{\G YM}_{\G M}
+n^{XYM}_{YM}-n^{XYM}_{XM}\notag\\
&=\frac1{2\pi}\Bigl[
 (\beta^{XM}-\beta^{\G M})|_{U_{\G XM}}
 -(\beta^{YM}-\beta^{\G M})|_{U_{\G YM}}
 +(\beta^{YM}-\beta^{XM})|_{U_{XYM}}\Bigr]\notag\\
&=\frac1{2\pi}\Bigl[
 (\beta^{XM}|_{U_{\G XM}}-\beta^{XM}|_{U_{XYM}})
 +(\beta^{\G M}|_{U_{\G YM}}-\beta^{\G M}|_{U_{\G XM}})\notag\\
&\hspace{35mm}
 +(\beta^{YM}|_{U_{XYM}}-\beta^{YM}|_{U_{\G YM}})\Bigr]\notag\\
&\equiv r_M[\beta]\pmod2.
\end{align}
Here $r_P[\beta]$ is the winding number on the quotient defined in the
main text, equal to half the winding number along a small circle around
$P$ in $T^2$.

Finally, we express the determinant parities in terms of the
one-dimensional invariants. Cover the interval from $P$ to $Q$ with real
frames $\Phi^P,\Phi^Q$ satisfying the symmetry conditions. Within each patch,
\begin{align}
\tr\bigl[(\Phi^A)^\dagger d\Phi^A\bigr]=0,
\qquad A=P,Q
\end{align}
so the determinant of the Wilson line reduces to $\det t^{PQ}$ at the
frame change. The gauge condition at the high-symmetry points makes the
time-reversal sewing matrix equal to the same standard antisymmetric
matrix $J$ at both endpoints, so the Pfaffian ratio is $1$.
Thus, the partial polarization gives
\begin{align}
e^{i\g_{\rm T}(P\to Q)}=\det t^{PQ}=(-1)^{p^{PQ}}.
\end{align}
In particular,
\begin{align}
p^{\G X}&=\nu^{C_2}_{1,\G\to X},&
p^{\G Y}&=\nu^{C_2}_{1,\G\to Y},&
p^{YM}&=\nu^{C_2}_{1,Y\to M}.
\label{eq:tree_edge_parities}
\end{align}
We thus obtain
\begin{align}
\nu_2^{C_2}[\alpha+\beta]-\nu_2^{C_2}[\alpha]
&=r_X[\beta]\nu^{C_2}_{1,\G\to X}
 +(r_Y[\beta]+r_M[\beta])\nu^{C_2}_{1,\G\to Y}
 +r_M[\beta]\nu^{C_2}_{1,Y\to M}\notag\\
&=r_X[\beta]\nu^{C_2}_{1,\G\to X}
 +r_Y[\beta]\nu^{C_2}_{1,\G\to Y}
 +r_M[\beta]\nu^{C_2}_{1,\G\to M}
\quad\bmod2.
\label{eq:alpha_general_change_appendix}
\end{align}
The last equality uses the additivity of partial polarization along the path $\G\to Y\to M$.

\bibliography{refs}
\end{document}